\documentclass[12pt,emlines2]{article}
\usepackage{amssymb}
\usepackage{amscd}
\usepackage{euscript}
\begin{document}
\setcounter{footnote}{0}
\begin{center}
\vspace{1.0in}
{\Large\bf Isomonodromic deformations of the sl(2) Fuchsian systems on the Riemann sphere}
\end{center}
\bigskip
\begin{center}
{\sc  Sergey OBLEZIN\footnote{\it e-mail: oblezin@mccme.ru, sergey.oblezin@itep.ru}} \\
\end{center}
\begin{center}
{\it Institute for Theoretical and Experimental Physics,\\
25, Bol. Cheremushkinskaya, Moscow, 117259.}\\
\vspace{2mm}

\end{center}
\bigskip
\abstract{\footnotesize This paper is devoted to two geometric
constructions related to the isomonodromic method. We follow
Drinfeld's ideas from \cite{D3} and develop them in the case of the
curve $X=\mathbb{P}^1\setminus\{a_1,\ldots,a_n\}$. Thus we
generalize the results of \cite{AL} to the case of arbitrary number
$n$ of points. First, we construct separated Darboux coordinated in
terms of the Hecke correspondences between moduli spaces. In this
way we present a geometric interpretation of the Sklyanin formulas
from \cite{Skl}. In the second part of the paper, we construct
Drinfeld's compactification of the initial data space and describe
the compactifying divisor in terms of certain FH-sheaves. Finally,
we give a geometric presentation of the dynamics of the
isomonodromic system in terms of deformations of the compactifying
divisor and explain the role of apparent singularities for Fuchsian
equations. To illustrate the results and methods, we give an example
of the simplest isomonodromic system with four marked points known
as the Painlev´e-VI system.
}\\
\bigskip

\vspace{2mm} \noindent Key words: isomonodromic method, separation
of variables, Drinfeld's compactification, the Frobenius-Hecke
sheaves, the Painlev\'e-VI equation.

\vspace{2mm} \noindent MSC: 15A54, 32G02, 34B02, 14E07



\section{Introduction}

\noindent  In this work we present a geometric interpretation of the
isomonodromic deformation of a non-resonant $sl(2)$ Fuchsian system
on the Riemann sphere;  this isomonodromic system is known as the
Schlesinger system. Given a generic Fuchsian differential equation
of order $N$ with singularities at $S:=\{a_1,\ldots,a_n\}$ on
$\mathbb{P}^1$ and let us put it into an isomonodromic analytical
family in the following way.

Consider the Fuchsian system of differential equations
${\displaystyle
  \frac{d}{dz}Y(z)=\left(\sum\limits_{i=1}^n\frac{B_i}{z-a_i}\right)Y(z)}$
with matrix coefficients $B_i\in{\it Mat}(N,\mathbb{C})$. Note from
the very beginning that we consider only non-resonant systems, that
is we assume that $\lambda_i^{(a)}-\lambda_i^{(b)}\notin\mathbb{N}$
for the eigenvalues $\{\lambda_i^{(a)}\}$ of matrices
$B_i,\,i=1,\ldots,n$. Let $Y(z)$ be the fundamental solution of this
system. Then consider the equation $\partial_{\it
z}Y(z)\,=L(z)Y(z),$ where
$$L(z)=\sum\limits_{i=1}^n\frac{B_i(a_1,\ldots,a_n)}{z - a_i}dz$$
simultaneously with the following (isomonodromic) condition for the
coefficients $B_i(a_1,\ldots,a_n)$, $i=1,\ldots,n$
$$dB_i(a_1,\ldots ,a_n)\,=\,\sum\limits_{j=1}^n\,[B_j, B_i]\,d\log(a_i-a_j)$$
called the Schlesinger equation. The last equation indicates the
complete integrability condition $d\omega=\omega\wedge\omega$ for
the matrix-valued 1-form $\omega=dY(z)\cdot Y(z)^{-1}$; in other
words, it is the zero-curvature condition for the logarithmic
connection $\nabla:=d-\omega$ in a trivial rank $N$ bundle on the
configuration space $\mathbb{P}^1\times\mathbb{C}^n$. Such systems
were investigated originally by Schlesinger (see \cite{Sch}) and
later algebraic aspects were considered by Flashka and Newell (see
\cite{FN}), Jimbo and Miwa (see \cite{JM}). Geometric aspects of the
isomonodromic systems were originally initiated by R\"ohrl in
\cite{R}, and then developed by Bolibruch (see \cite{B}, \cite{AB}),
Hitchin (see \cite{Hit}), Arinkin and Lysenko (\cite{AL}) in various
senses. In our approach we develop Drinfeld's ideas (see \cite{D2},
\cite{D3}) to study the isomonodromy problem from the point of view
of geometric representation theory; in particular, we generalize the
results of \cite{AL} to the case of arbitrary number of
singularities.

In this paper we describe the fundamental matrix of our Fuchsian
system of rank 2 in terms of horizontal sections of a certain rank 2
bundle $\mathcal{L}$ with respect to the logarithmic
$sl(2)$-connection on the Riemann sphere $\mathbb{P}^1$
$$\nabla:\,\mathcal{L}\longrightarrow
\mathcal{L}\otimes\Omega_{\mathbb{P}^1}^1(a_1+\ldots+a_n)$$ with
${\rm Res}_{a_i}\nabla=B_i$.

Consider the co-adjoint representation of the group
$G=SL(2)\rightarrow{\rm End}(sl(2)^*),\,X\mapsto {\rm ad}_X^*$;
herewith we assume that the coefficients $B_i$ lie in co-adjoint
$sl(2)$-orbits $\EuScript{O}_i$. Fixing the eigenvalues of the
residues $B_i$ we fix the appropriate $sl(2)$-orbits. Every
$sl(2,\mathbb{C})$-orbit is a 2-dimensional non-compact variety with
a natural symplectic form which in the co-adjoint representation is
$\omega_\xi(X,Y)=-\langle{\rm ad}_X^*,Y\rangle$ for any $\xi\in
sl(2)^*$. The symplectic quotient of the direct product of
$SL(2,\mathbb{C})$-orbits is a symplectic variety of the following
dimension
$${\rm dim}\Big(\prod\limits_{i=1}^n\EuScript{O}_i
//SL(2,\mathbb{C})\Big)
=n\cdot{\rm dim}\,\EuScript{O}_i-2\cdot {\rm
dim}\,SL(2,\mathbb{C})=2(n-3).$$ Identify the symplectic quotient
with the initial data space and present it as an open subset of the
coarse moduli space $\mathcal{M}_n(2)$ of collections
$$(\mathcal{L},\nabla;\quad\phi: {\rm Det}\mathcal{L}\simeq
\mathcal{O}_{\mathbb{P}^1};\quad\lambda_1,...,\lambda_{\it n} ),$$
of a rank 2 bundle $\mathcal{L}$ on $\mathbb{P}^1$ with fixed
determinant and a connection
$\nabla:\mathcal{L}\rightarrow\mathcal{L}\otimes\Omega^1_{\mathbb{P}^1}
(a_1+\ldots+a_n)$ such that the eigenvalues of the residues of the
connection ${\rm Res}_{a_i}\nabla=B_i$ at ${\it a_i},\,i=1,\ldots
,n$ are $\{\lambda_i,\,-\lambda_i\}$. Note that in this work we
consider only the coarse moduli spaces and assume this even when the
word "coarse" is omitted.

In his remarkable thesis \cite{D2} Drinfeld introduced original
geometric objects; they are elliptic modules and the Frobenius-Hecke
sheaves. In our paper we follow the geometric ideas of Drinfeld and
apply them to investigate the isomonodromic deformation of the
Fuschsian systems of rank two. We pay a special attention to the
bounds of application for the procedure of separation of variables.
In particular, we construct Drinfeld's compactification (see
\cite{D2}) of the initial data space of the system and investigate
the cases in which the classical procedure of separation of
variables does not work. Also emphasize that our construction
naturally entails the identification of the phase space of a
Fuchsian system of differential equations (initial data space) with
the phase space of the isomonodromic deformation of the system. In
this way one may understand this paper as a geometric presentation
of the \emph{isomonodromic method} of investigation of Fuchsian
differential equations.

The paper is organized in the following way. In the first part
(Section 3) we construct geometric Darboux coordinates on
$\mathcal{M}_n(2)$ and notice that the result coincides with the
calculations ("magic recipe") from \cite{Skl}.

Let us point that we omit the assumption of the triviality of the
bundle $\mathcal{L}$ though fixing its determinant by a horizontal
isomorphism $\phi:{\rm Det}\mathcal{L}\simeq\mathcal{O}$. We
consider the moduli space $\mathcal{M}_n(2)$ of pairs
$(\mathcal{L},\nabla)$ equipped with $\phi$ and with fixed
eigenvalues of the residues of the connection. We construct a
parametrization of the moduli space in the sense of Drinfeld and in
this way we give a geometric interpretation of the Sklyanin's
formulas from \cite{Skl}. For this purposes we have to impose a
notion of \emph{stability} for our configurations. We discuss it and
investigate the isomonodromic system for the (semi)stable
configurations. Note, that addition of the strata of
$\mathcal{M}_n(2)$ which correspond to non-trivial bundles allows to
uncover the hidden symmetries of the system; in particular, the
discrete symmetries of the isomonodromic system , calculated in
\cite{O}, can be explained only in terms of the completed initial
data space $\mathcal{M}_n(2)$ (see the example in the end of this
paper).

The second part of the paper (Section 4) contains the construction
of Drinfeld's compactification of the initial data space
$\mathcal{M}_n(2)$. Below we present two naive recipes how to
complete the initial data space of the isomonodromic deformation.
They are very simple and explicit; however, they are both
generalized by our construction of the compactification and
illustrate it.

Consider the cotangent bundle $\Omega$ on $\mathbb{P}^1$ and denote
by ${\rm Tot}(\mathbb{P}^1,\Omega(\sum a_i))$ the total space of the
bundle $\Omega(a_1+\ldots+a_n)$. Our construction of the Darboux
coordinates provides the description of the initial data space
$\mathcal{M}_n(2)$ in terms of the $(n-3)$-th symmetric power of the
non-compact surface $K_n:={\rm Tot}(\mathbb{P}^1,\Omega(\sum a_i))$
(see for example \cite{GNR}). Precisely, consider the compact
surface $\overline{K_n}=\mathbb{P}(\mathcal{O}\oplus\Omega(\sum
a_i))=s_\infty\sqcup K_n$ for $s_\infty$ the infinite section
$s_\infty$ and let $F_i:=\Omega(\sum
a_i)|_{a_i}\subset\overline{K_n}$. The fibres $F_i,\,i=1,\ldots,n$
are trivialized by the residue map
$R:\,F_i\widetilde{\rightarrow}\mathbb{C}$. Let us blow-up the
surface $\overline{K_n}$ at $2n$ points $R^{-1}(\lambda^\pm_i)\in
F_i$ for
$\{\lambda^\pm_i\}=\{\lambda_1,1-\lambda_1,\,\lambda_2,-\lambda_2,\ldots,
\lambda_n,-\lambda_n\}$ and consider the non-compact surface
$$K'_n:=({\rm Bl}_{R^{-1}(\pm\lambda_i)}\overline{K_n})\setminus
(s_\infty\cup\widetilde{F_1}\cup\ldots\cup\widetilde{F_n}),$$
where $\widetilde{F_i}$ are the proper pre-images of the fibers
$F_i,\,i=1,\ldots,n$. There is a map
$\mathcal{M}_n(2)\longrightarrow(K'_n)^{(n-3)}:=(K'_n)^{n-3}/
\mathfrak{S}_{n-3}$
that is an isomorphism at the generic point and we thoroughly
describe points where it is not an isomorphism.

There is no an ordering on the set of the variables
$\{x_i,p_i\},\,i=1,\ldots,n-3$ and hence we have to consider either the
quotient $(K'_n)^{(n-3)}:=(K'_n)^{n-3}/\mathfrak{S}_{n-3}$ or the
$(n-3)!$-covering
$$\widetilde{\mathcal{M}_n(2)}\simeq(K'_n)^{n-3}.$$ On the covering
$\widetilde{\mathcal{M}_n(2)}$ we have a natural symplectic form
$$\varpi=\sum\limits_{i=1}^{n-3}dx_i\wedge dp_i$$ and it equips
$\widetilde{\mathcal{M}_n(2)}$ with a structure of symplectic
variety. It is natural to complete $\mathcal{M}_n(2)$ with a
pole-divisor of the symplectic form $\varpi$. We present the
compactifying divisor as the pole-divisor of $\varpi$ in 5.2.

Another natural way to regard the compactification problem is as
follows. Consider an algebraic curve $C$ on $K_n$ defined by the
equation $R(z,\lambda)=0$ for
$$R(z,\lambda):=({\rm det}L(z)-\lambda\cdot{\it Id}).$$ It is known
as \emph{spectral curve}; the genus of $C$ is $n-3$, which is equal
to the half of dimension of the initial data space. Remarkably, $C$
is not preserved by the isomonodromic deformation and this fact
entails the natural completion of the phase space with a limit cycle
of the spectral curve $C$; this construction of completion of the
initial data space is a part of the isomonodromic method.

It is significant that it is possible to perform a natural
compactification of the initial data space $\mathcal{M}_n(2)$ in
terms of a degenerated model of the curve $C$. Consider the surface
$\overline{K_n}$, trivialized fibers $F_i,\,i=1,\ldots,n$ and the
infinite section $s_\infty$ on it; let $\{ F_i,\,s_\infty\}$ be the
basis in the homology group $H_2(\overline{K_n},\mathbb{Z})$. The
intersection numbers are
$$F_i\cdot F_j=0,\quad s_\infty\cdot s_\infty=-{\rm deg}\,\Omega(a_1+\ldots+a_n)=2-n,
\quad F_i\cdot s_\infty=1,\quad C\cdot F_i=2;$$ besides, the
intersection number of the curve $C$ with $s_\infty$ is zero. The
topological class of $C$ is preserved by the isomonodromy
deformation. In this way we compactify the initial data space
$\mathcal{M}_n(2)$ with the divisor $D$ such that its factors
$\Theta_{(i)}\subset\overline{K'_n}$ preserve the topological
invariant and $\Theta_{(i)}\cdot s_\infty=0$; for $n=4$ this
immediately implies $\Theta=2s_\infty+F_1+\ldots+F_4$. In the case
$n>4$ the above argument is not so explicit and we obtain the same
result in 4.2 using the FH-sheaves approach to the compactification
problem.

Besides, in the fourth section we emphasize the important role of
the complete self-intersection of the compactifying divisor: ${\bf
\Theta}_n:=D\cdot D$ whose dimension is exactly $n-3$. We describe
the dynamics of the isomonodromic system in terms of the cycle ${\bf
\Theta}_n$. Finally we explain the role of the apparent
singularities of the Fuchsian systems originally introduced in
\cite{F}(see also \cite{B} and \cite{AB}). Precisely, we identify
the cycle ${\bf\Theta}_n$ with the moduli space of the collections
$$(\widetilde{\mathcal{L}_{\Theta_n}},\,\nabla_{\Theta_n};\,
\phi':{\rm Det}\,\widetilde{\mathcal{L}_{\Theta_n}}
\widetilde{\rightarrow}\mathcal{O}(-a_1-\ldots-a_{n-2});\,
(\widetilde{\lambda_i^+},\,\widetilde{\lambda_i^-}),\,i=1,\ldots,n)$$
for some $a\in S$, where $\widetilde{\mathcal{L}_{\Theta_n}}$ is the
rank 2 bundle of degree $2-n$ with the logarithmic connection
$\nabla_{\Theta_n}$ such that the eigenvalues of {\rm Res}$_{\it
a_i}\nabla_{\Theta_n}$ are
$(\widetilde{\lambda_1^+},\widetilde{\lambda_1^-}):=
(\lambda_i,1-\lambda_i)$ at $a_i,\,i=1,\ldots,n-2$ and
$(\widetilde{\lambda_i^+}, \widetilde{\lambda_i^-}):=(\lambda_i,
-\lambda_i)$ at $a_i= a_{n-1},\,a_n$. We present the dynamical
variables $\{x_i,p_i\},\,i=1,\ldots,n-3$ of the isomonodromic
deformation as the parameters of the Hecke correspondence between
${\bf\Theta}_n$ and the moduli space
$\mathcal{M}'_n(2)\simeq\mathcal{M}_n(2)$ of the collections
$$(\widetilde{\mathcal{L}},\widetilde{\nabla}:=
\nabla|_{\widetilde{\mathcal{L}}};\widetilde{\phi}: {\rm
Det}\,\widetilde{\mathcal{L}}\simeq\mathcal{O}(-a_1); (\lambda_1^+,
\lambda_1^-),\ldots ,(\lambda_n^+, \lambda_n^-)),$$ where
$\widetilde{\mathcal{L}}$ is a rank 2 bundle on $\mathbb{P}^{1}$
with fixed horizontal isomorphism $\widetilde{\phi}:{\rm
Det}\mathcal{L}\simeq\mathcal{O}(-a_1)$ and with a connection
$\widetilde{\nabla}$ with singularities at $\{a_1,\ldots,a_n\}$; the
eigenvalues of {\rm Res}$_{\it a_i}\widetilde{\nabla}$ are
$(\lambda_1^+, \lambda_1^-):=(\lambda_1, 1-\lambda_1)$ at $a_1$ and
$(\lambda_i^+, \lambda_i^-):=(\lambda_i, -\lambda_i)$ at
$a_i,\,i=2,\ldots ,n$. In terms of the connections
$$\widetilde{\nabla}=\nabla_{\Theta_n}(p_1,\ldots,p_{n-3})-
\sum\limits_{i=1}^{n-3}{\bf P}_{p_i}\frac{dz}{z-x_i},$$ where
${\bf P}_{p_i}$ are the projectors on the invariant
one-dimensional subspaces
$p_i\subset\widetilde{\mathcal{L}_{\Theta_n}}|_{x_i},\,i=1,\ldots,n-3$.
The terms  ${\displaystyle {\bf P}_{p_i}\frac{dz}{z-x_i}}$ do not
change the monodromy of the connection and the points
$x_1,\ldots,x_{n-3}$ are called \emph{the apparent singularities}
of the connection $\widetilde{\nabla}$.

\subsection{Acknowledgements}

\noindent I am deeply grateful to my Ph.D. advisor A. Levin for
numerous fruitful stimulating discussions, in particular, for
teaching me the FH-sheaves technique and for discussions of the D.
Arinkin and S. Lysenko papers. I'm thankful to A. Zotov for useful
discussions. I appreciate A. Borodin, I. Krichever, and M.
Olshanetsky for their interest to this work. I am thankful to V.
Radionov for reading the text and numerous corrections of the
language. The work was also partially supported by the CRDF grant
RM1-2545, by the program for support of the scientific schools
NSh-1999.2003.2 and by the RFBR grant 04-01-00642.

\section{Modificaitons of logarithmic sl(2)-connections}

\noindent In \cite{D1} Drinfeld presented a construction of elliptic
module which generalized a set of classical algebraic ideas; then in
\cite{D2} the Frobenius-Hecke sheaves, (or, "shtukas") were
introduced. These new concepts provided a new understanding of the
Langlands conjecture for automorphic forms, and led to establishing
this conjecture in the case $GL(2)$ over function field. Besides,
this approach uncovered profound relations between arithmetic and
algebraic geometry, representation theory and differential
equations.

For our purposes it will be convenient to modify the original
definition from \cite{D2} and to introduce the following.\\
{\bf Definition.} A Frobenius-Hecke sheaf (FH-sheaf) of level $K$
(for an integer $K$) on $\mathbb{P}^1$ is a flag of locally free
sheaves of the same rank $\mathcal{F}_0\subset\mathcal{F}$ on
$\mathbb{P}^1$ such that the codimension of the support ${\rm
supp}\,(\mathcal{F}/\mathcal{F}_0)\subset\mathbb{P}^1$ equals one
and $(\mathcal{F}/\mathcal{F}_0)$ has a $K$-dimensional space of
sections. For a generic FH-sheaf all the points of ${\rm
supp}\,(\mathcal{F}/\mathcal{F}_0)$ are distinct that is
$\mathcal{F}/\mathcal{F}_0$ is isomorphic to a sum of sky-scraper
sheaves $\bigoplus\delta_{x_i}$ and each sky-scraper sheaf
$\delta_{x_i}$ has a one-dimensional space of
sections.\\
Between the moduli spaces of FH-sheaves
$(\mathcal{F}'_1\subset\mathcal{F}_1)$ and
$(\mathcal{F}'_2\subset\mathcal{F}_2)$ of different levels $K_1$ and
$K_2$ there are correspondences, called the Hecke correspondences.
These correspondences are performed by modifications (see \cite{D3})
of the locally free sheaves $\mathcal{F}'_i,\,\mathcal{F}_i$; upper
modifications reduce the level and lower ones increase it.

Given a rank 2 bundle $\mathcal{L}$ on $\mathbb{P}^{1}$ with a
connection $\nabla$, let ${\it x} \in \mathbb{P}^{1}$. Denote by $V$
a fiber $\mathcal{L}|_{\it x}$ and let ${\it l}\subset{\it V}$ be a
one-dimensional subspace. Identify $\mathcal{L}$ with of its sheaf
sections and consider the following locally trivial sheaves.
$$({\it x},{\it l})^{\rm low}(\mathcal{L}):=\{{\it s}\in\mathcal{L} ~|
~{\it s(x)}~\in{\it l}\},\qquad ({\it x, l})^{\rm up}(\mathcal{L}):=
({\it x, l})^{\rm low}(\mathcal{L})\otimes\mathcal{O}({\it x})$$
which are called the lower and the upper modifications respectively.
Denote the lower modification by $\widetilde{\mathcal{L}}:=({\it
x},{\it l})^{\rm low}(\mathcal{L})$ and consider the natural map
$\widetilde{\mathcal{L}}|_{\it x} \longrightarrow\mathcal{L}|_{\it
x}$; evidently its image is {\it l}. Set $\widetilde{l}:= {\rm
ker}(\widetilde{\mathcal{L}}|_{\it x}\longrightarrow
\mathcal{L}|_{\it x})$ then $({\it x}, \widetilde{l})^{\rm
up}\widetilde{\mathcal{L}}~=~\mathcal{L}.$ The lower and the upper
modifications provide the following exact sequences.
$${\rm 0}\longrightarrow({\it x, l})^{\rm low}(\mathcal{L})
\longrightarrow\mathcal{L}\longrightarrow\delta_{\it x}\otimes
\mathcal{L}_{\it x}/{\it l}\longrightarrow ~{\rm 0},$$
$${\rm 0}\longrightarrow\mathcal{L}\longrightarrow({\it x, l})^{\rm up}
\mathcal{L}\longrightarrow\delta_{\it x}\otimes {\it
l}\otimes\mathcal{T}_{\it x}\longrightarrow{\rm 0}$$ respectively.
Here $\delta_{\it x}$ is a sky-scraper sheaf with the support at
{\it x} and $\mathcal{T}_x$ is the localization of the tangent
bundle at $x$.

Roughly speaking, given a local decomposition ${\it V}\,=\,{\it
l}\bigoplus\widetilde{\it l}$ of $\mathcal{L}\simeq{\it
V}\otimes\mathcal{O}$, we have
$$({\it x, l})^{\rm low}(\mathcal{L})={\it l}\otimes\mathcal{O}\bigoplus
\widetilde{l}\otimes\mathcal{O}({\it -x}),\qquad ({\it x, l})^{\rm
up}(\mathcal{L})={\it l}\otimes\mathcal{O}({\it x})
\bigoplus\widetilde{l}\otimes\mathcal{O}.$$

In other words we change our bundle rescalling the basis of sections
in the neighborhood of a point {\it x}; if the local basis is
$\{{\it s}_1({\it z}),{\it s}_2({\it z})\}$ with ${\it
l}\otimes\mathcal{O}\simeq \langle{\it s}_1({\it z})\rangle$ and
${\widetilde{\it l}}\otimes\mathcal{O}\simeq \langle{\it s}_2({\it
z})\rangle$ then the basis of the lower modification $(x, l)^{\rm
low}$ of the bundle is generated by the sections $\{{\it
s}_1(z),\,({\it z-x}){\it s}_2(z)\},$ and of the upper one $(x,
l)^{\rm up}$ by $\{({\it z-x})^{-1}\,{\it s}_1(z),\,{\it s}_2(z)\}.$
Consequently, in the punctured neighborhood we may represent the
action of the modifications by the following gluing matrices.
$$({\it x, l})^{\rm low}=
\left(\begin{array}{cc}
 1 & 0 \\
0 & ({\it z-x})
\end{array}\right),\qquad
({\it x, l})^{\rm up}= \left(\begin{array}{cc}
({\it z-x})^{-1} & 0 \\
0 & 1
\end{array}\right).$$
Matrix presentation of the modifications is supposed to be quite
obvious, and further on we use it freely. Let us note that in our
setting the discussed Hecke correspondences are symplectic
(singular) gauge transformations (see \cite{LOZ}).

Now discuss the action of the modifications of an $sl(2)$-connection
with
logarithmic singularities on the projective line $\mathbb{P}^1$.\\
{\bf Definition. \cite{S}} A \emph{modulus} $\mathfrak{M}$ supported
at {\it S} on an algebraic curve $X$ is a finite set
$S=\{a_1,...,a_n\}\subset X$ equipped with a function assigning a
positive integer ${\it n_i}$ to every point ${\it a_i}\in S$.
Sometimes we identify $\mathfrak{M}$ with the effective divisor
$\sum n_i\cdot a_i\,$. In the present work we consider the module
$$\mathfrak{M}\,=\,\sum\limits_{i=1}^n\,a_i.$$

Let us look how the modifications change the connection. Suppose we
start from some logarithmic (Fuchsian) $sl(2)$-connection $\nabla$
on $\mathcal{L}$ and
$$\nabla:\,\mathcal{L}\longrightarrow\mathcal{L}\otimes\Omega^{\rm 1}
(\mathfrak{M});$$ this means that $\nabla$ has \emph{simple poles}
at the support $S$ of $\mathfrak{M}$. Denote the eigen-subspaces of
{\rm Res}$_{a_i}\nabla$ by $\ell_i^{\pm}:={\rm ker}({\rm
Res}_{a_i}\nabla \mp \lambda _i)$ and consider the modifications of
our pair $(\mathcal{L}, \nabla)$ in these subspaces. Emphasize that
we modify the pairs $(\mathcal{L},\,\nabla)$ in (${\rm Res}_{\it
x}\nabla$)-invariant subspaces of ${\it V}\subseteq\mathcal{L}|_{\it
x}$, otherwise we increase the order of a pole of the connection.
Indeed, using the matrix presentation write down the action of the
modification of the bundle in a non-invariant subspace at $z=0$:
$$
\left(\begin{array}{cccc}
1 & &  0 \\
\\
0 & & z
\end{array}\right)
\left[{\it d}\quad +\quad \left(\begin{array}{cc}
\lambda/z & \varepsilon/z \\
\\
0 & -\lambda/z
\end{array}\right)
\right]
\left(\begin{array}{cc}
1 & 0 \\
\\
0 & 1/z
\end{array}\right)
~=~{\it d} +
\left(\begin{array}{cc}
\lambda/z & \varepsilon/z^2 \\
\\
0 & -(\lambda+1)/z
\end{array}\right),$$ where $z$ is a local parameter. Here
because of the $\varepsilon$ in the right upper corner, the second
component of the modification is not $\nabla$-invariant.

Besides, note that the lower and upper modifications at any point
${\it x}~\in\mathbb{P}^{1}$ change the determinant:
$${\rm Det}({\it x, l})^{\rm low}\mathcal{L} =
{\rm Det}\mathcal{L}\otimes\mathcal{O}(-x),\quad {\rm Det}({\it x,
l})^{\rm up}\mathcal{L} ={\rm Det}\mathcal{L}\otimes
\mathcal{O}(x).$$

Let us illustrate the techniques that we will use in the next
sections. Consider the lower modification $\widetilde{\mathcal{L}}$
with the connection
$$\nabla': \widetilde{\mathcal{L}}
\stackrel{\nabla|_{\widetilde{\mathcal{L}}}}{\longrightarrow}
\mathcal{L}\otimes\Omega(\mathfrak{M}) \stackrel{{\rm
pr}}{\longrightarrow}
\widetilde{\mathcal{L}}\otimes\Omega(\mathfrak{M})$$ on
$\widetilde{\mathcal{L}}$ then on the determinant bundle we get the
connection
$${\rm Tr}\nabla' = {\rm Tr}\nabla +\frac{dz}{z-x}.$$

Perform a pair of the lower and the upper modifications at points
$a_i$ and $a_j$ respectively to get the bundle $\mathcal{L}''$
with the same determinant
$${\rm Det}\mathcal{L}''={\rm Det}\mathcal{L}\otimes
\mathcal{O}({\it a}_{\it j} - {\it a}_{\it i}) \simeq{\rm
Det}\mathcal{L};$$ to do this we have to fix a set of compatible
isomorphisms $\mathcal{O}\simeq\mathcal{O}({\it a_i-a_j})$ such that
$$\mathcal{O}\simeq\mathcal{O}({\it
a_i-a_j})\otimes\mathcal{O}({\it a}_{\it j}-{\it a}_{\it k})\simeq
\mathcal{O}({\it a}_{\it i}-{\it a}_{\it k}).$$

Nevertheless, if we start from an $sl(2)$-connection $\nabla$, then
after such procedure we get the connection
$$\nabla'' = \nabla + {\bf P}_{\it l_i}\frac{dz}{z - a_i} -
{\bf P}_{\widetilde{l}_{\it j}}\frac{dz}{z - a_j},$$ where ${\bf
P}_*$ are the projectors on the appropriate {\rm
Res}$\nabla$-invariant subspaces; it is the $gl(2)$-connection. In
order to get $sl(2)$-connection we have to add the suitable 1-form
$$\widetilde{\nabla}''=\nabla''+\frac{1}{2}\left({\bf 1}_2
\frac{dz}{z-a_j}-{\bf 1}_2\frac{dz}{z-a_i}\right),$$
where ${\bf 1}_2$ denotes the identity $2\times 2$ matrix.

For two points ${\it a_i}, {\it a_j}\in S$ consider the modified
$SL(2)$-bundle
$$\mathcal{L}'' = ({\it a}_{\it j}, l_{\it j}^{+})^{\rm up}\circ
({\it a}_{\it i}, l_{\it i}^{-})^{\rm low}\mathcal{L}$$ with
modified logarithmic connection $\nabla''$ defined above. This
provides a nontrivial transformations of the coarse moduli space
$\mathcal{M}_{\it n}$ of rank 2 bundles with fixed horizontal
isomorphism $\phi$ and logarithmic connection with fixed eigenvalues
of residues on $\mathbb{P}^{1}$; in other words we have the Hecke
correspondence on $\mathcal{M}_n$ as follows.
\\
{\bf Proposition. (\cite{O})} The modified pair ($\mathcal{L}'',
\widetilde{\nabla''}$) is an element of the coarse moduli space
$\mathcal{M}_{\it n}$. The eigenvalues of {\rm
Res}$_{a}\widetilde{\nabla''},\,a\in S$ are
$$\{\lambda_1,\ldots,\lambda_{\it i}+\frac{1}{2},\ldots,
\lambda_{\it j}-\frac{1}{2},\ldots,\lambda_{\it n}\}$$ for the case
of a pair of modifications at distinct points $a_i, a_j\in{\it S}$;
for a pair of modifications at one point $a_k\in S$, the eigenvalues
are
$$\{\lambda_1,\ldots,\lambda_{\it k}+1,\ldots,\lambda_{\it n}\}.$$

In this way, we have birational isomorphisms between the moduli
spaces with different parameters, or between different initial data
spaces; the group structure is isomorphic to the affine Weyl group
$\mathfrak{W}(\widehat{C_n})$. For precise description of the
discrete symmetries of our system and their action on the local
solutions see \cite{O}.

\section{Separation of variables}

\noindent In this section, following \cite{AL}, we describe our
initial data and construct \'etale coordinates on the open subset of
$\mathcal{M}_n(2)$. It appears that these coordinates are separated
coordinated in the sense of Sklyanin. Originally the recipe for the
separation of variables was introduced in \cite{FMcL} for the
periodic Toda model. Then this procedure was generalized to the case
of the Gaudin model by Sklyanin (\cite{Skl}). Our calculation of
separated variables in terms of $\Omega(\mathfrak{M})$-valued
operator $L(z)$ coincides with Sklyanin's "magic recipe". In this
way we give a geometric interpretation of the Sklyanin's separation
of variables for the Gaudin model.

We generalize the results of the Arinkin and Lysenko work \cite{AL}
and present the calculations for an arbitrary number $n$ of
singularities; however, we use the ideas from \cite{AL}, in
particular, two linear-algebraic lemmas.

Fix a collection $\lambda_1,\ldots ,\lambda_n$ of complex numbers
and the modulus $\mathfrak{M}$ with the support $S$ at distinct
points $a_ 1,...,a_n$ on $\mathbb{P}^1$. The group of projective
automorphisms of the Riemann sphere being three-dimensional, it is
natural to restrict ourselves to the case of $n\geq 3$. Suppose
$\mathcal{L}$ be a rank 2 bundle on $\mathbb{P}^{1}$ equipped with a
fixed horizontal isomorphism $\phi:{\rm
Det}\mathcal{L}\simeq\mathcal{O}$ and a connection $\nabla$ with
singularities at $\mathfrak{M}=\sum{\it a_i}$; the eigenvalues of
{\rm Res}$_{\it a_i}\nabla$ are $(\lambda_i,
-\lambda_i),\,i=1,\ldots ,n$.

\subsection{Stable bundles}
\noindent Let us discuss the definition of \emph{stability} of our
data. We consider the moduli space of vector bundles of rank 2 and
we permanently control the pair $(\mathcal{L},\nabla)$ to be
indecomposable in order to provide the stability. For these purposes
we put the following eigenvalue condition
$$\sum\epsilon_{\it i}\lambda_{\it i}\notin\mathbb{Z},\qquad
(\epsilon_1, \ldots ,\epsilon_n)\in(\mathbb{Z}/{\rm
2}\mathbb{Z})^n,$$ which guarantees the irreducibility of the pair
"bundle $\mathcal{L}$ with the connection $\nabla$" and implies the
stability of this pair. Indeed, given a $\nabla$-invariant rank 1
sub-bundle $\mathcal{L}_1\subset\mathcal{L}$ equipped with a
connection $\nabla_1:=\nabla|_{\mathcal{L}_1}$ then
$(\mathcal{L}_1)|_{a_i}\subset\mathcal{L}|_{a_i}$ is an eigen-space
of ${\rm Res}_{a_i}\nabla$ and ${\rm Res}_{a_i}\nabla_1$ is an
eigenvalue of ${\rm Res}_{a_i}\nabla$. In this way we get ${\rm
Res}_{a_i}\nabla_1=\pm\lambda _i$ but from the other hand $\sum{\rm
Res}_{a_i}\nabla_1=-{\rm deg}\,\mathcal{L}_1\in\mathbb{Z}$
contradicts our eigenvalue-condition.

Moreover, our bundle $\mathcal{L}$ with the trivial determinant is
in general nontrivial and may have a structure
$\mathcal{O}(k)\oplus\mathcal{O}(-k)$. The value of $k$ depends on
$n$ and it is defined by the stability of the construction in the
following way. Let $\mathcal{L}_0:=\mathcal{O}(k)$ be a sub-bundle
then by irreducibility we have a non-zero map
$$\nabla_0:\mathcal{L}_0\rightarrow (\mathcal{L}/\mathcal{L}_0)\otimes
\Omega(\mathfrak{M})$$ which implies
$${\rm deg}\,\mathcal{L}_0\leq{\rm deg}(\mathcal{L}/\mathcal{L}_0)+
{\rm deg}\,\Omega(\mathfrak{M}) =0-{\rm
deg}\,\mathcal{L}_0+n-2,\quad\mbox{hence,}\quad
k\leq\frac{n-2}{2}.$$ We consider the moduli space of pairs
$(\mathcal{L},\nabla)$ and look after the automorphism group of the
pair. We demand ${\it Aut}(\mathcal{L},\nabla)=\mathbb{C}^*$ and we
assume that there are no $\nabla$-invariant sub-bundles
$\mathcal{L}_0\subset\mathcal{L}$.

\subsection{The map
$(\mathcal{L},\nabla)\mapsto(\mathcal{L}_0\subset\mathcal{L},\nabla)$}

\noindent We shall act in the following way. Suppose that we can
choose a \emph{distinguished} sub-bundle
$\mathcal{L}_0\subset\mathcal{L}$. Then we will investigate the
features of a (semi)stable element
$(\mathcal{L},\nabla)\in\mathcal{M}_n(2)$ looking at its restriction
on the (non-invariant) \emph{distinguished} sub-bundle. We have seen
that for $(\mathcal{L},\nabla)\in\mathcal{M}_n$ the structure of our
bundle $\mathcal{L}$ can be $\mathcal{O}(k)\oplus\mathcal{O}(-k)$
for some $k$ but, for example, if $k=0$ and
$\mathcal{L}\simeq\mathcal{O}\oplus\mathcal{O}$ then there is no way
to choose the distinguished sub-bundle. The fact is that a bundle of
an odd degree always has a distinguished sub-bundle, and it is in
this way that we have to modify our bundle.

Take a point from $S$, say, $a_1$ and consider the bundle
$\widetilde{\mathcal{L}}:=(a_1, l^+_1)^{\rm low}\mathcal{L}$. The
natural embedding $\widetilde{\mathcal{L}}\subset\mathcal{L}$
provides an isomorphism $\mathcal{M}_n(2)\simeq\mathcal{M}_n'(2)$
with the moduli space of the following collections.

$$(\widetilde{\mathcal{L}},
\widetilde{\nabla}:=\nabla|_{\widetilde{\mathcal{L}}};\widetilde{\phi}:
{\rm
Det}\widetilde{\mathcal{L}}\simeq\mathcal{O}(-a_1);(\lambda_1^+,
\lambda_1^-), \ldots ,(\lambda_n^+, -\lambda_n^-)).$$ Here
$\widetilde{\mathcal{L}}$ is a rank 2 bundle on $\mathbb{P}^{1}$
with a fixed horizontal isomorphism $\widetilde{\phi}:{\it
det}\mathcal{L}\simeq\mathcal{O}(-a_1)$ and with a logarithmic
connection $\widetilde{\nabla}$ with singularities at
$\{a_1,\ldots,a_n\}$. The eigenvalues of {\it Res}$_{\it
a_i}\widetilde{\nabla}$ are $(\lambda_1^+, \lambda_1^-):=(\lambda_1,
1-\lambda_1)$ at $a_1$ and $(\lambda_i^+, \lambda_i^-):=(\lambda_i,
-\lambda_i)$ at $a_i,\,i=2,\ldots ,n$. The dimension of the vector
space of embeddings
$\mathcal{L}/\mathcal{L}_0\simeq\mathcal{O}(-k)\hookrightarrow\mathcal{L}$
for $k>0$ equals
$${\rm dim}\,{\rm Hom}(\mathcal{O}(-k),\mathcal{O}(k))=2k+1=3,\ldots
,2\cdot\left[\frac{n-2}{2}\right]+1.$$ Thus, we can choose a
sub-bundle $\mathcal{O}(-k)$ passing through at least $2k+1$ of $n$
lines $l_i^+:={\rm ker}({\rm Res}_{x_i}-\lambda _i)$ and then at
least one line lies neither in $\mathcal{L}_0$, nor in our chosen
$\mathcal{O}(-k)$, as we assume the bundle
$(\mathcal{L};\phi;l_i,\,i=1,\ldots ,n)$ to be irreducible. Thus we
get the distinguished sub-bundle
$\widetilde{\mathcal{L}_0}\subset\widetilde{\mathcal{L}}$ with
possible values of degree ${\rm
deg}\widetilde{\mathcal{L}_0}:=k'=0,\ldots ,[\frac{n-2}{2}]$. For
example, in both cases $n=4$ and $n=5$ the structure of
$\mathcal{L}$ can be only $\mathcal{O}\oplus\mathcal{O}$ and
$\mathcal{O}(1)\oplus\mathcal{O}(-1)$; nevertheless for $n=4$ the
modified bundle is always
$\widetilde{\mathcal{L}}\simeq\mathcal{O}\oplus\mathcal{O}(-1)$ and
for $n=5$ it can be either $\mathcal{O}\oplus\mathcal{O}(-1)$, or
$\mathcal{O}(1)\oplus\mathcal{O}(-2)$, since the direction of the
modification $l_1^+$ can lie in $\mathcal{L}_0\simeq\mathcal{O}(1)$.

\subsection{$\mathcal{M}'_n(2)$ as a moduli space of FH-sheaves}

\noindent The algebraic variety $\mathcal{M}_n\simeq\mathcal{M}_n'$
is non-compact and consists of locally closed strata
$\EuScript{M}^{k'}$, which can be interpreted as the moduli space of
the following collections.
$$(\mathcal{O}(k')\subset\widetilde{\mathcal{L}};\widetilde{\nabla};
\widetilde{\phi}:{\rm Det}\widetilde{\mathcal{L}}\simeq
\mathcal{O}(-a_1);\,(\lambda^+_1,\lambda^-_1),\ldots,
(\lambda^+_n,\lambda^-_n))$$ indexed by $k'$. The maximal value of
$k'$ depends on the parity of $n$: if $n$ is even, then
$k'=\frac{n-4}{2}$, and if $n$ is odd, then $k'=\frac{n-3}{2}$.

Pick a collection of points $y_1,\ldots ,y_{k'}\in\mathbb{P}^1$, and
fix an isomorphism
$\widetilde{\mathcal{L}_0}\simeq\mathcal{O}(y_1+\ldots +y_{k'})$;
then choose a connection $\nabla_0$ with respect to this isomorphism
with $k'$ simple poles precisely at $y_1,\ldots ,y_{k'}$ such that
$$\nabla_0:\widetilde{\mathcal{L}_0}\longrightarrow
\widetilde{\mathcal{L}_0}\otimes\Omega(y_1+\ldots +y_{k'}),\qquad
{\rm Res}_{y_i}\nabla_0=1.$$ Fixing the connection $\nabla_0$ we get
a distinguished trivialization (section)
$\mathcal{O}\hookrightarrow\widetilde{\mathcal{L}_0}$ of our
sub-bundle.

Restrict the connection on the sub-bundle
$\widetilde{\mathcal{L}_0}$ and consider the map
$$B:=\widetilde{\nabla}|_{\widetilde{\mathcal{L}_0}}-\nabla_0:\quad
\widetilde{\mathcal{L}_0}\rightarrow
\widetilde{\mathcal{L}}\otimes\Omega(\mathfrak{M}).$$ In this way
we obtain the maps
$$f_{k'}:\EuScript{M}^{k'}\rightarrow M_1:=\,\mbox{moduli space of}\,
(\widetilde{\mathcal{L}_0}\simeq
\mathcal{O}(k')\subset\widetilde{\mathcal{L}},B),$$ where
$\widetilde{\mathcal{L}}/\widetilde{\mathcal{L}_0}\simeq\mathcal{O}(-k'-1)$ and
$B:\mathcal{T}(-\mathfrak{M})\hookrightarrow\widetilde{\mathcal{L}}$
for $\mathcal{T}(-\mathfrak{M}):=\Omega(\mathfrak{M})^{-1}$.

Using the maps $f_{k'}$ we construct the maps from our moduli space
$\mathcal{M}_n'$ to the moduli space of the so-called Drinfeld
FH-sheaves (see \cite{D2}):
$$\{\mathcal{O}\oplus\mathcal{T}(-\mathfrak{M})\subset
\widetilde{\mathcal{L}}|\,\widetilde{\mathcal{L}}/(\mathcal{O}\oplus
\mathcal{T}(-\mathfrak{M}))\simeq\Delta_{n-3}\},$$ where ${\rm
dim}\Gamma(\mathbb{P}^1,\Delta_{n-3})=n-3$ and ${\it
supp}(\Delta_{n-3})$ is in codomension one.

To present the strata of $\mathcal{M}'_n(2)$ as moduli spaces we
have to reconstruct the element
$(\widetilde{\mathcal{L}},\widetilde{\nabla})\in\mathcal{M}'_n(2)$
from the FH-sheaf
$A=(\mathcal{O}\oplus\mathcal{T}(-\mathfrak{M})\subset
\widetilde{\mathcal{L}})$.\\
{\bf Proposition. \cite{AL}} Let $A$ be an FH-sheaf of level $n-3$
and let $R_i$ be an operator
$\widetilde{\mathcal{L}}|_{a_i}\rightarrow\widetilde{\mathcal{L}}|_{a_i}$
with eigenvalues $\lambda_i^\pm$. Then, on the stratum
$\EuScript{M}^0$ there is a
unique connection $\widetilde{\nabla}$ such that in the above notations\\
(i) $\widetilde{\nabla}|_{\widetilde{\mathcal{L}}_0}=d+B$ for the
unique connection
$d:\widetilde{\mathcal{L}}_0\rightarrow\widetilde{\mathcal{L}}_0\otimes
\Omega$
the unique connection;\\
(ii) ${\it Res}_{a_i}\nabla=R_i;$\\
(iii) $(\widetilde{\mathcal{L}},\widetilde{\nabla})\in\mathcal{M}'_n(2).$

In this way we identify the generic stratum $\EuScript{M}^0$ with
the moduli space of certain FH-sheaves. On the other strata the
connection $\widetilde{\nabla}$ is not unique and in the following
two subsections we prove the analogous proposition for all the
strata. In the next subsections we give a simple construction from
linear algebra and calculate the affine space of connections
$\widetilde{\nabla}$ that satisfy conditions (i)-(iii).

\subsection{A construction from the linear algebra}

\noindent In terms of the linear algebra our description of stable
pairs $(\widetilde{\mathcal{L}},\widetilde{\nabla})$ is nothing but
a reconstruction of the operator $L(z)$ such that
$(\widetilde{\mathcal{L}},\partial_z-L(z))\in\widetilde{\mathcal{M}_n'}$
from the first row $B$ of the operator $L$ and the eigenvalues of
the residues. Let $V_0\subset V\simeq\mathbb{C}^2$ be a complete
flag of vector spaces and let $R_0\in{\rm Hom}(V_0,V)$.
\\
{\bf Lemma A. \cite{AL}} Let $\lambda^+\neq\lambda^-\in\mathbb{C}$
and put $\mathbf{R}:=\{R\in{\rm End}(V)$ such that $R|_{V_0}=R_0$
and the eigenvalues of $R$ are $\lambda^+,\lambda^-\}$,
$\mathbf{L}:=\{(l^+\neq l^-)|\,l^\pm\subset V,\,{\rm dim}\,l^\pm =1$
with $(R_0-\lambda^\mp)(V_0)\subset l^\pm\}$. Then the map
$$F:\mathbf{R}\longrightarrow\mathbf{L},\quad R\mapsto
({\rm ker}(R-\lambda^+)=
{\rm im}(R-\lambda^-),{\rm ker}(R-\lambda^-))$$ is bijective.\\
{\it Proof.} Clearly, $F$ is injective, so let us check the
surjectivity. For $(l^+,l^-)\in \mathbf{L}$ denote the corresponding
projectors by $P_\pm :V\rightarrow V/l^\pm\simeq l^\mp$; one has
$P_++P_-={\rm Id}$. The condition $(R_0-\lambda^\mp)(V_0)\subset
l^\pm$ implies $P^\mp(R_0-\lambda^\mp)(V_0)=0$, or,
$P^-(R_0-\lambda^-)(V_0)+P^+(R_0-\lambda^+)(V_0)=0$; hence,
$R_0=(\lambda^+P^++\lambda^-P^-)|_{V_0}$ and for
$R:=(\lambda^+P^++\lambda^-P^-)\in\mathbf{R}$ we have
$F(R)=(l^+,l^-)$.
$\,\blacksquare$\\
One can make the similar calculations for the case $l^+=l^-$ and proof
the analogous statement.
\\
{\bf Lemma B.} Let $\lambda:=\lambda^+=\lambda^-\in\mathbb{C}$ and
put $\mathbf{R}:=\{R\in{\rm End}(V)$ such that $R|_{V_0}=R_0$ and
$R$ has the only one eigenvalue $\lambda\}$, $\mathbf{L}:=\{(l\neq
l')|\,l,l'\subset V,\, {\rm dim}\,l,l' =1$ with
$(R_0-\lambda)(V_0)\subset l$ and $(R_0-\lambda)(l')\subset V_0\}$.
Then the map
$$F:\mathbf{R}\longrightarrow\mathbf{L},\quad R\mapsto({\rm ker}
(R-\lambda),{\rm im}(R-\lambda))$$ is bijective.$\,\blacksquare$

\subsection{Calculation of the affine space of connections}

\noindent In this subsection we use the notations and technique from
\cite{SGA5}. Let us remark that the connection $\widetilde{\nabla}$
that satisfies conditions (i)-(iii) exist locally on $\mathbb{P}^1$.
Given an open subset $U\subset\mathbb{P}^1$, denote by
$\mathcal{C}(U)$ the set of all local connections
$\widetilde{\nabla} =\nabla_0 -L(z)$ on $U$. Given
$\widetilde{\nabla},\widetilde{\nabla}'\in\mathcal{C}(U)$, then
$E:=\widetilde{\nabla}-\widetilde{\nabla}'$ is an element of
$H^0(U,\underline{\it Hom}
(\widetilde{\mathcal{L}},\widetilde{\mathcal{L}}\otimes\Omega))\simeq
\underline{\it Hom}
(\widetilde{\mathcal{L}}/\widetilde{\mathcal{L}_0},
\widetilde{\mathcal{L}_0}\otimes\Omega)$ such that
$E|_{\widetilde{\mathcal{L}_0}}=0$ and ${\rm Tr}E=0$. Denote by
$\mathcal{E}(U)$ the set of such local homomorphisms. Clearly,
$\mathcal{C}$ is an $\mathcal{E}$-torsor and the obstruction to the
existence of a global connection lies in
$H^1(\mathbb{P}^1,\mathcal{E}(\mathfrak{M}))$ which by the Serre
duality is dual to $$H^0(\mathbb{P}^1,\,\{ E\in {\rm
End}(\mathcal{L})\,|\,{\rm Tr}\,E=0,\,E(a_i)(l^+_i)\subset
l^+_i\,\})=\{0\}.$$ In this way a global connection always exists,
but it is not unique.

Thus we parameterize the space of connections by the matrix element
$L(z)_{21}$ of $L(z)$, and as we have seen
$$L(z)_{21}\in\underline{\it
Hom}(\widetilde{\mathcal{L}}/\widetilde{\mathcal{L}_0},
\widetilde{\mathcal{L}_0}\otimes\Omega)\simeq\mathcal{E}.$$ Let us
calculate the space of connections on each stratum
$\EuScript{M}^{k'}$, assuming that the FH-sheaf
$A=(\mathcal{O}\oplus\mathcal{T}(-\mathfrak{M})\subset
\widetilde{\mathcal{L}})$ is generic.

\emph{On the stratum $\EuScript{M}^0$} we have the following diagram
$$
\begin{CD}
0 @>>> \mathcal{O}\oplus\mathcal{T}(-\mathfrak{M}) @>A>>
\widetilde{\mathcal{L}} @>>> \bigoplus_{i=1}^{n-3}\delta_{x_i}\otimes
p_i\otimes\mathcal{T}_{x_i} @>>> 0\\
@. @| @| @. @.\\
0 @>>> \mathcal{O}\oplus\mathcal{O}(2-n) @>A>>
\mathcal{O}\oplus\mathcal{O}(-1)
\end{CD}$$
For all $x_i$ we have ${\rm
im}\,A(x_i)\nsubseteq\widetilde{\mathcal{L}_0}\simeq\mathcal{O}$,
hence, all $p_i<\infty$ and the map
$$\EuScript{M}^0\longrightarrow
\underbrace{K_n'\times\ldots\times K_n'}_{n-3}$$ is an isomorphism
at a generic point (modulo the assumption that all $x_i$ are
distinct). The sheaf $\mathcal{E}\simeq\underline{\it Hom}
(\widetilde{\mathcal{L}}/\widetilde{\mathcal{L}_0},
\widetilde{\mathcal{L}_0}\otimes\Omega)$ is of degree $-1$, hence,
any $\mathcal{E}$-torsor is trivial and we have the unique
connection recovered by our procedure.

\emph{On the stratum $\EuScript{M}^1$} we have
$$A:={\rm Id}\oplus B:\quad\mathcal{O}\oplus(\mathcal{T}(-\mathfrak{M}))
\longrightarrow
\widetilde{\mathcal{L}}\simeq\mathcal{O}(y_1)\oplus\mathcal{O}(-2)$$
and, if $x_i=y_1$ for some $i$, then we make the upper modification
at $x_i$ in the infinite direction, and $p_i=\infty$. Note that the
case $p_i=\infty$ corresponds to the point at infinity of
$\overline{K_n'}:=\mathbb{P}(\mathcal{O}\oplus\Omega(\mathfrak{M}))$
and it means that the modification in
$(\mathcal{O}\oplus\mathcal{T}(-\mathfrak{M}))|_{x_i}$ is performed
in the direction of $\mathcal{O}|_{x_i}\subset(\mathcal{O}\oplus
\mathcal{T}(-\mathfrak{M}))|_{x_i}$. In this way we have a map
$$\EuScript{M}^1\longrightarrow\overline{K_n'}\times
\underbrace{K_n'\times\ldots\times K_n'}_{n-4}$$ The sheaf
$\mathcal{E}=\underline{\it Hom}
(\widetilde{\mathcal{L}}/\widetilde{\mathcal{L}_0},
\widetilde{\mathcal{L}_0}\otimes\Omega)$ is isomorphic to
$\underline{\it Hom}(\mathcal{O}(-2),\mathcal{O}(1)\otimes\Omega)
\simeq\mathcal{O}(1)$ and on this stratum the affine space of
connections is 2-dimensional.

\emph{On the stratum $\EuScript{M}^{k'}$} we have
$$A:={\it Id}\oplus B:\quad\mathcal{O}\oplus(\mathcal{T}(-\mathfrak{M}))
\longrightarrow\widetilde{\mathcal{L}}\simeq\mathcal{O}(y_1+\ldots
+y_{k'})\oplus\mathcal{O}(-k'-1),$$ hence,
$$\EuScript{M}^{k'}\longrightarrow\underbrace{\overline{K_n'}\times\ldots
\times\overline{K_n'}}_{k'}\times\underbrace{\overline{K_n'}\times\ldots
\times\overline{K_n'}}_{n-3-k'}.$$ Besides,
$\mathcal{E}\simeq\underline{\it
Hom}(\mathcal{O}(-k'-1),\mathcal{O}(k')\otimes\Omega)
\simeq\mathcal{O}(2k'-1)$, and on this stratum the affine space of
connections is parameterized by $L(z)_{21}$, and it is
$2k'$-dimensional.

\subsection{\'Etale coordinates on $\mathcal{M}'_n(2)$ at the generic point}

\noindent Recall that from
$$L|_{\widetilde{\mathcal{L}_0}}=B:\quad\mathcal{T}(-\mathfrak{M})
\hookrightarrow \widetilde{\mathcal{L}}$$ and ${\rm
Id}:\mathcal{O}\hookrightarrow \widetilde{\mathcal{L}}$ we have
constructed FH-sheaf
$$A:={\rm Id}\oplus B:\quad\mathcal{O}\oplus\mathcal{T}(-\mathfrak{M})
\longrightarrow \widetilde{\mathcal{L}}.$$ Moreover, in the generic
situation we have the following factorization
$$A=A_1\circ\ldots\circ A_{n-3},\quad A_i=(x_i,p_i)^{\rm up},\,i=1,\ldots,n-3$$ which implies
${\rm Det}A(x_i)=0,\,i=1,\ldots,n-3$; hence, in the neighborhood of
a point $x_i$ we have
$$A(x_i)=\left(
\begin{array}{cc}
B_{11} & B_{12}\\
1 & 0
\end{array}\right)\quad\mbox{and}\quad
B_{11}(x_i)=p_i,\quad B_{12}(x_i)=0,\,i=1,\ldots ,n-3.$$ By Lemmas A
and B we recover the operator $L(z)$ from the following data;
$L(z)|_{\mathcal{L}_0}=A(z)$, ${\rm Res}_{a_i}L(z)$ has the
eigenvalues $\lambda_i^+,\lambda_i^-$ and the trace ${\rm
Tr}L(z)=(z-a_1)^{-1}$.

The $n-3$ zeroes of $B_{12}$ are exactly the $x_i,\,i=1,\ldots ,n-3$
\'etale coordinates on $\mathcal{M}'_n$. One readily identify this
calculation with the analogous one from \cite{Skl}.

In this way we are given an exact sequence
$$0\longrightarrow\mathcal{O}\oplus\mathcal{T}(-\mathfrak{M})\stackrel{A}
{\longrightarrow}\widetilde{\mathcal{L}}\longrightarrow
\delta_{x_i}\otimes p_i\otimes\mathcal{T}_{x_i}\longrightarrow
0,\qquad i=1,\ldots,n-3,$$ where $A_1\circ\ldots\circ
A_{n-3}=A:\mathcal{O}\oplus\mathcal{T}(-\mathfrak{M})\rightarrow
\widetilde{\mathcal{L}}$ is a composition of the upper modifications
$(x_i,p_i)^{\rm up}$. The directions of the modifications
$p_i\subset(\mathcal{O}\oplus\mathcal{T}(-\mathfrak{M}))|_{x_i}$ are
one-dimensional subspaces and they are parameterized by the surface
${\rm Tot}(\mathbb{P}^1,\Omega(\mathfrak{M}))$. So, we would like to
construct maps $\mathcal{M}'_n(2)\longrightarrow{\rm
Tot}(\mathbb{P}^1,\Omega(\mathfrak{M}))$ and parameterize
$\mathcal{M}'_n(2)$ by $\{x_i,p_i\},\,i=1,\ldots ,n-3$. In fact
$\{x_i,p_i\},\,i=1,\ldots,n-3$ are \'etale coordinates on an open
subset of $\mathcal{M}'_n(2)$.

There is no ordering on our array of $A_i,\,i=1,\ldots,n-3$ and we
have the action of the symmetric group $\mathfrak{S}_{n-3}$ on our
construction of $\mathcal{M}'_n$; a change of order of the upper
modifications $A_i=(x_i,p_i)^{\rm up},\,i=1,\ldots,n-3$ induces a
nontrivial automorphism of ${\rm
Tot}(\mathbb{P}^1,\Omega(\mathfrak{M}))^{n-3}$. In this way, there
is no a map from $\mathcal{M}'_n$ to ${\rm
Tot}(\mathbb{P}^1,\Omega(\mathfrak{M}))^{n-3}$, but there is one to
the quotient
$${\rm Tot}(\mathbb{P}^1,\Omega(\mathfrak{M}))^{(n-3)}:=\underbrace{
{\rm Tot}(\mathbb{P}^1,\Omega(\mathfrak{M}))\times\ldots\times {\rm
Tot}(\mathbb{P}^1,\Omega(\mathfrak{M}))}_{n-3}/\mathfrak{S}_{n-3}.$$
One may also consider the $(n-3)!$-branched covering
$\widetilde{\mathcal{M}'_n}$ of $\mathcal{M}'_n$, and study the
interplay between $\widetilde{\mathcal{M}'_n}$ and ${\rm
Tot}(\mathbb{P}^1,\Omega(\mathfrak{M}))^{n-3}$.

\subsection{Description of the fibers $F_i=\Omega({\mathfrak{M}})|_{a_i}$}

\noindent Let us analyze the behavior of the map $A$ when $x_i$
tends to $a\in S$. At a singular point $a$ we have two conditions
foon the eigen-values of the residue $L_a:={\rm Res}_a\nabla$:
$${\rm Tr}\,L_a=0\quad\mbox{and}\quad{\rm Det}\,L_a=\lambda_a
^+\cdot\lambda_a ^-,\quad a\in S.$$ We reconstruct the operator
$$L(z)|_{x_i\rightarrow a}=
\left(
\begin{array}{cc}
L_{11} & L_{12}\\
L_{21} & -L_{11}
\end{array}\right)$$
and obtain
$$L_{11}=B_{11}=p_idz,\quad {\rm Res}L_{12}\rightarrow 0,
\quad {\rm Res}L_{21}=\frac{{\rm Det}\,L_a-p_i^2}{{\rm
Res}L_{12}}.$$ We see that ${\rm Res}\,L(z)_{21}$ can have a finite
value only when $p_i\rightarrow\lambda_a ^\pm$ and we have to
calculate the limit of $L_{21}$ by the L'Hospital rule considering
the next terms of expansions of ${\rm Det}\,L_a-p_i^2$ and ${\rm
Res}L_{12}$. From the geometric point of view we just make a blow-up
(a $\sigma$-process) at this point.

Consider $K_n:={\rm
Tot}(\mathbb{P}^1,\mathcal{O}\oplus\Omega(\mathfrak{M}))$ with the
fibers $F_a\subset K_n$ at $a\in\mathbb{P}^1$. Since ${\rm
Res}_a:\Omega(\mathfrak{M})|_a\stackrel{\sim}{\rightarrow}\mathbb{C}$,
we have $R_a:F_a\stackrel{\sim}{\rightarrow}\mathbb{C}$; blow up
$K_n$ at $2n$ points $R_a^{-1}(\lambda_a ^\pm)$ and get
$$K_n':=({\rm Bl}_{R_a^{-1}(\lambda_a ^\pm)}K_n)\setminus\bigsqcup
\widetilde{F_a},$$ where $\widetilde{F_a}$ are the pre-images of
the fibers $F_a\subset K_n$ after the blow-up processes. Finally,
we have a map
$$\widetilde{\mathcal{M}'_n}\longrightarrow\underbrace{K_n'\times\ldots
\times K_n'}_{n-3}.$$ For $n=4$ this map is an isomorphism but, in
general as we have seen in 3.5 this map is neither injective nor
surjective; nevertheless, it is an isomorphism at the generic point
of $\widetilde{\mathcal{M}'_n}$.

\section{Compactification and dynamics of the system}

\noindent We have found the \'etale coordinates
$\{x_i,p_i\},\,i=1,\ldots,n-3$ on the open subset of the initial
data space $\mathcal{M}'_n(2)$ and now we investigate a
compactification of $\mathcal{M}'_n(2)$ in terms of these variables.
On the open subset of the moduli space $\mathcal{M}_n(2)$ is
isomorphic to the symmetric power of the surface $K'_n$; each factor
is $(K'_n)_{(i)}\simeq{\rm
Bl}_{\lambda_i^\pm}\mathbb{P}(\mathcal{O}\oplus
\Omega(\mathfrak{M}))\setminus\Theta_{(i)},\,i=1,\ldots,n-3$. In the
same way the factors of the compactifying divisor $D$ are the
components
$$(\Theta_{(i)})^{\rm red}=
s_\infty+\widetilde{F_1}+\ldots+\widetilde{F_n}\subset
\mathcal{B}l_{\lambda_i^\pm}\mathbb{P}(\mathcal{O}\oplus
\Omega(\mathfrak{M})),$$ where $s_\infty$ is the infinite section
$\mathbb{P}(\mathcal{O}\oplus\Omega^1(\mathfrak{M}))\setminus {\it
Tot}(\Omega^1(\mathfrak{M})$ and $\widetilde{F_i}$ are the
pre-images of the fibres
$F_i:=\Omega^1(\mathfrak{M})|_{a_i}\subset{\rm
Tot}(\Omega^1(\mathfrak{M})$ at singular points $a_1,\ldots,a_n$. In
this way the compactifying divisor is
$$D={\bf\Theta}_n+\sum_{r=1}^{n-3}(\Theta_{(i)})^r\times(K'_n)^{n-3-r},$$
where ${\bf\Theta}_n=D\cdot D$ is the complete self-intersection
cycle and evidently ${\bf\Theta}_n=\Theta^{(n-3)}.$

In this section we present a natural compactification of
$\mathcal{M}'_n(2)$ due to Drinfeld (see \cite{D2}). Namely, we use
the interpretation of $\mathcal{M}'_n(2)$ as moduli space of
FH-sheaves with ceratin restricting conditions. Thus to complete
such moduli space one just has to remove the restricting conditions
on FH-sheaves. Moreover, this construction gives a description of
the compactifying set as a moduli space of certain FH-sheaves. At
the end of the section in 4.3 we give a geometric presentation of
isomonodromic dynamics in terms of ${\bf \Theta}_n$.

\subsection{Drinfeld's compactification}

\noindent Note that all the moduli spaces considered here are the
coarse moduli spaces, and we do not discuss here the interplay
between the corresponding algebraic stacks. Recall the
interpretation of the moduli space $\mathcal{M}_n(2)$ in terms of
certain FH-sheaves step by step.

First, we present an isomorphism
$\mathcal{M}_n(2)\stackrel{\sim}{\rightarrow}\mathcal{M}'_n(2)$,
where $\mathcal{M}'_n(2)$ is the moduli space of rank 2 bundles
$\widetilde{\mathcal{L}}$ with the horizontal isomorphism
$\widetilde{\phi}:{\rm
Det}\widetilde{\mathcal{L}}\simeq\mathcal{O}(-a_1)$. This bundle is
equipped with a logarithmic connection $\widetilde{\nabla}$ with
fixed eigenvalues $\{\lambda_i^+,\lambda^-_i\}$ of the residues
${\rm  Res}_{a_i}\nabla$. This isomorphism is given by the lower
modification $\widetilde{\mathcal{L}}:=(a_1,l_1^+)^{\rm low}$ in the
direction $l_1^+:={\rm ker}({\rm Res}_{a_1}\nabla - \lambda_1)
\subset\widetilde{\mathcal{L}}|_{a_1}$ and the eigenvalues of the
residues of the connection are
$$\lambda_i^+=\lambda_i,\,i=1,\ldots,n,\qquad
\lambda_1^-=1-\lambda_1,\quad\lambda_i ^-=-\lambda_i,\,i>1.$$ The
upper modification $(a_1,l_1^-)^{\rm up}$ defines the inverse
isomorphism.

Second, the pair $(\widetilde{\mathcal{L}},\nabla)$ is irreducible
and contains a distinguished sub-sheaf
$\widetilde{\mathcal{L}_0}\subset\widetilde{\mathcal{L}}$ of
degree
  $k'=0,\ldots,\left[\frac{n-3}{2}\right]$. We fix a set of distinct
  points $y_1,\ldots,y_{k'}\in\mathbb{P}^1$ such that
$$\widetilde{\mathcal{L}_0}\stackrel{\sim}{\rightarrow}
\mathcal{O}(y_1+\ldots+y_{k'})$$ and consider a connection
$$\nabla_0:\,\widetilde{\mathcal{L}_0}\longrightarrow
\widetilde{\mathcal{L}_0}\otimes\Omega(y_1+\ldots+y_{k'});$$
fixing $\nabla_0$ we define a distinguished section
$\mathcal{O}\subseteq\widetilde{\mathcal{L}_0}$.

Denote by $M_1$ the coarse moduli space of triples
$$(\widetilde{\mathcal{L}_0}\subset\widetilde{\mathcal{L}},\,A,\,
\widetilde{\phi}),$$
where
$$\widetilde{\mathcal{L}}/\widetilde{\mathcal{L}_0}\simeq
\mathcal{O}(-k'-1),\quad k'=0,\ldots,\left[\frac{n-3}{2}\right],$$
and $A\in{\rm
Hom}(\widetilde{\mathcal{L}_0},\widetilde{\mathcal{L}}\otimes
\Omega(\mathfrak{M}))$ such that ${\rm
im}(A)\nsubseteq\widetilde{\mathcal{L}_0}\otimes\Omega(\mathfrak{M})$.
There is a map $\mathcal{M}'_n(2)\rightarrow M_1,$ defined by
$$(\widetilde{\mathcal{L}},\nabla,\widetilde{\phi})\,\mapsto\,
(\widetilde{\mathcal{L}_0}\subset\widetilde{\mathcal{L}},\,
A:=\nabla|_{\widetilde{\mathcal{L}_0}}-\nabla_0,\,\widetilde{\phi}).$$
Note that on the open subset the moduli space $M_1$ is isomorphic to
the $(n-3)$-th symmetric power of the non-compact surface ${\rm
Tot}(\mathbb{P}^1,\Omega(\mathfrak{M}))$ and the condition ${\rm
im}(A)\subset\widetilde{\mathcal{L}_0}\otimes\Omega(\mathfrak{M})$
defines the infinite section $s_\infty\subset{\rm
Tot}(\mathbb{P}^1,\Omega(\mathfrak{M}))$.

Third, identify the moduli space $\mathcal{M}'_n(2)$ with the
coarse moduli space of the following collections;
$$(\widetilde{\mathcal{L}_0}\subset\widetilde{\mathcal{L}},\,A,\,
\widetilde{\phi};\,l_1^+,l_1^-,\ldots,l_n^+,l_n^-),$$
such that\\
(i) $(\widetilde{\mathcal{L}_0}\subset\widetilde{\mathcal{L}},\,A,\,
\widetilde{\phi})$
is a point of the moduli space $M_1$;\\
(ii) $l_i^\pm\subset\widetilde{\mathcal{L}}|_{a_i}$ is the
one-dimensional subspace defined by
$$({\rm Res}_{a_i}A-\lambda^\mp)(\widetilde{\mathcal{L}_0}|_{a_i})
\subset l_i ^\pm;$$
(iii) $l_i^+\neq l_i^-$.\\
In the previous section it was shown that on the open subset we may
identify the $(n-3)!$-covering $\widetilde{\mathcal{M}_n(2)}$ with
the $(n-3)$-th power of the surface $K'_n$. The surface
$K'_n\simeq{\rm Bl}_{\lambda ^\pm}{\rm
Tot}(\mathbb{P}^1,\Omega(\mathfrak{M}))$ is obtained by blowing up
$K_n={\rm Tot}(\mathbb{P}^1,\Omega(\mathfrak{M}))$ at $2n$ points
$(a_i,\lambda ^\pm)$.

Denote by $M_2$ the coarse moduli space of
$(\widetilde{\mathcal{L}_0}\subset\widetilde{\mathcal{L}},\,A,\,
\widetilde{\phi};\,l_1^+,l_1^-,\ldots,l_n^+,l_n^-)$ such that only
the conditions (i), (ii) are satisfied, and (iii) is hold for all
$a_i$ except for some $a\in S$. It is the condition (iii) that
defines the union of pre-images of the fibers
$F_i:=\Omega^1(\mathfrak{M})|_{a_i}\subset{\rm
  Tot}(\Omega^1(\mathfrak{M}))$ and the infinite section $s_\infty$.
Thus $M_2$ is a divisor on $\mathcal{M}'_n(2)$; moreover, it
naturally completes our moduli space $\mathcal{M}'_n(2)$ and we
identify $M_2$ with the compactifying divisor $D$. It is the
Drinfeld compactification in the sense of \cite{D2}.

Denote by $M'_2$ the coarse moduli space of
$(\widetilde{\mathcal{L}_0}\subset\widetilde{\mathcal{L}},\,A,\,
\widetilde{\phi};\,l_1^+,l_1^-,\ldots,l_n^+,l_n^-)$ such that only
the conditions (i), (ii) are satisfied; the condition (iii) does not
hold for all $a_i\in S$. Identify $M'_2$ with the complete
self-intersection locus of the compactifying divisor $D$ and denote
it by ${\bf \Theta}_n$.

\subsection{$D$ and ${\bf \Theta}_n$ in terms of FH-sheaves}

As we have seen the divisor $D$ (and its complete self-intersection
${\bf\Theta}_n$) may be identified with the coarse moduli space of
$(\widetilde{\mathcal{L}_0}\subset
\widetilde{\mathcal{L}},\,A,\,\widetilde{\phi})$ with $A\in{\rm
Hom}(\widetilde{\mathcal{L}_0},\widetilde{\mathcal{L}}
\otimes\Omega(\mathfrak{M}))$,
satisfied the following two conditions:\\
(1) ${\rm im}(A)\subset\widetilde{\mathcal{L}_0}\otimes
\Omega(\mathfrak{M})$;\\
(2) $l_a^-:=({\rm Res}_{a}A-\lambda^+)
(\widetilde{\mathcal{L}_0}|_{a})=l_i^+:=({\rm Res}_{a}A-\lambda^-)
(\widetilde{\mathcal{L}_0}|_{a})$ for some (and for all) $a\in S$.\\
Condition (2) implies
$l_a^+=l_a^-=(\widetilde{\mathcal{L}_0}|_{a})$, and for $a=a_i$ it
defines the fibre $F_i$. Altogether, conditions (2) imply (1), and
the (1) means that all the subspaces $l_i^+$ and $l_i^-$, for
$i=1,\ldots,n$, coincide with $\widetilde{\mathcal{L}_0}|_{a_i}$ and
define the (blow-up of the) intersection of all fibers
$F_i,\,i=1,\ldots,n$. In this way the conditions (1) and (2) give us
components
$$\Theta_{(i)}:=(s_\infty+s_\infty+F_1+\ldots+F_n)\subset\overline{K_n}.$$

Consider the Hecke correspondence between our moduli space
${\bf\Theta}_n$ of FH-sheaves
$(\mathcal{O}\oplus\mathcal{T}(-\mathfrak{M})\subset
\widetilde{\mathcal{L}})$ of level $n-3$ and the moduli space
${\bf\Theta}'_n$ of FH-sheaves
$(\mathcal{O}\oplus\mathcal{T}(-\mathfrak{M})\subset
\widetilde{\mathcal{L}}')$ of level zero. In other words, let us
perform $n-3$ lower modifications of our bundle
$\widetilde{\mathcal{L}}$ of degree -1 at distinct points
$a\in\{a_1,\ldots,a_n\}$ in the direction $l_a ^+=l_a ^-$. Thus,
after such procedure we get the bundle
$\mathcal{O}\oplus\mathcal{T}(-\mathfrak{M})$ of degree $2-n$ for
the chosen directions $l_a ^+=l_a ^-$ lie in
$\widetilde{\mathcal{L}_0}|_{a_i}$.

It is more convenient to investigate the complete self-intersection
locus ${\bf\Theta}_n$ of the compactifying divisor
$D=\overline{\mathcal{M}'_n(2)}\setminus\mathcal{M}'_n(2)$. In fact,
it is isomorphic to the coarse moduli space of collections
$$(\widetilde{\mathcal{L}_{\Theta_n}},\,\nabla_{\Theta_n},\,\phi'),$$
with the fixed eigenvalues of residues of the connection. Here
$\widetilde{\mathcal{L}_{\Theta_n}}$ is a bundle of degree $2-n$ on
$\mathbb{P}^1$ with the horizontal isomorphism $\phi':{\rm
Det}\widetilde{\mathcal{L}}'
\stackrel{\sim}{\rightarrow}\mathcal{O}(-a_1-a_{n-2})$, and the
connection $\nabla_{\Theta_n}$ has the following eigenvalues of the
residues. For $a_i=a_1,\ldots,a_{n-2}$ the residues ${\rm
Res}\nabla_{\Theta_n}$ have eigenvalues $(\lambda_i,1-\lambda_i)$
and for $a_i= a_{n-1},\,a_n$ the eigenvalues are
$(\lambda_i,-\lambda_i)$.

The connection $\nabla_{\Theta_n}$ exists but it is not unique. Let
us calculate the dimension of the appropriate affine space. Given an
open subset $U\subset\mathbb{P}^1$, denote by $\mathcal{C}(U)$ the
set of all local connections $\nabla_{\Theta_n} =\nabla_0-L(z)$ on
$U$. For two connections
$\nabla_{\Theta_n}',\nabla_{\Theta_n}''\in\mathcal{C}(U)$ their
difference $E':=\nabla_{\Theta_n}''-\nabla_{\Theta_n}'$ is an
element of $H^0(U,\underline{\it Hom}
(\widetilde{\mathcal{L}_{\Theta_n}},\widetilde{\mathcal{L}_{\Theta_n}}
\otimes\Omega))\simeq\underline{\it Hom}
(\widetilde{\mathcal{L}_{\Theta_n}}/\mathcal{O},\mathcal{O}\otimes\Omega)$,
such that $E'|_{\mathcal{O}}=0$ and ${\rm Tr}E'=0$. Let
$\mathcal{E}_{\Theta_n}(U)$ be the set of such morphisms $E'$. Then
$\mathcal{C}$ has a natural structure of
$\mathcal{E}_{\Theta_n}$-torsor and the obstruction to the existence
of a global connection lies in the group
$H^1(\mathbb{P}^1,\mathcal{E}_{\Theta_n}(\mathfrak{M}))$, which is
dual to $H^0(\{E'\in{\rm End}(\mathcal{L}_{\Theta_n})|{\rm
Tr}E'={\rm 0},\,E'(a_i)(l^+_i)\subset l^+_i\})=\{0\}$ by the Serre
duality. We define our global connection by reconstructing the row
$(L(z)_{21},-L(z)_{11})$ of the operator $L(z)$ and the connection
is parameterized by the element $L(z)_{21}$ that lies in
 $\underline{\it Hom}(\widetilde{\mathcal{L}_{\Theta_n}}/\mathcal{O},
\mathcal{O}\otimes\Omega)\simeq\mathcal{E}_{\Theta_n}$. In this way
$$\mathcal{E}_{\Theta_n}\simeq\Omega_{\mathbb{P}^1}^{\otimes 2}
(\mathfrak{M})\simeq\mathcal{O}(n-4)$$ and the dimension of the
affine space of the connection $\nabla_{\Theta_n}$ on the bundle
$\widetilde{\mathcal{L}_{\Theta_n}}\simeq\mathcal{O}\oplus\mathcal{T}
(-\mathfrak{M})$ equals $n-3$.

At last, just note that one can interpret the divisor $D$ as a
moduli space of certain FH-sheaves of level zero considering the
appropriate Hecke correspondence.

\subsection{Dynamics of the $sl(2)$ isomonodromic system}

\noindent In the final part of the section let us present the
\'etale coordinates $\{x_i,p_i\},\,i=1,\ldots,n-3$ as parameters of
the Hecke correspondence between the coarse moduli spaces
${\bf\Theta}_n$ and $\mathcal{M}'_n(2)$, and interpret them in terms
of the apparent singularities of the connection $\nabla$. Precisely,
consider the space of sections of the sheaf
$\mathcal{E}_{\Theta_n}\simeq\underline{\it
Hom}(\widetilde{\mathcal{L}_{\Theta_n}}/\mathcal{O},
\mathcal{O}\otimes\Omega ^1_{\mathbb{P}^1})$ on the moduli space
${\bf\Theta}'_n$ of the collections
$$(\widetilde{\mathcal{L}_{\Theta_n}},\,\nabla_{\Theta_n};\,
\phi':{\rm Det}\widetilde{\mathcal{L}_{\Theta_n}}
\stackrel{\sim}{\rightarrow}
\mathcal{O}(-a_1-\ldots-a_{n-2});\,(\widetilde{\lambda_i^+},\,
\widetilde{\lambda_i^-}),\,i=1,\ldots,n)$$ for
$\widetilde{\lambda_i^+}:=\lambda_i$ and
$\widetilde{\lambda_i^-}=1-\lambda_i\,$, for $a_i\neq
a_{n-1},\,a_n$; the rest $\widetilde{\lambda_i^-}=-\lambda_i$ for
$a_i= a_{n-1},\,a_n$. Note here that the configuration
$(\widetilde{\mathcal{L}_{\Theta_n}};\,l^+_1,\ldots,l^+_n)$ is
semi-stable in our notation, since we have
$${\rm Aut}(\widetilde{\mathcal{L}_{\Theta_n}})\simeq
\begin{array}{ccc}
{\it End}(\mathcal{O}) & \oplus & {\rm
Hom}(\mathcal{T}(-\mathfrak{M}),\,
\mathcal{O})\\
\oplus & \quad & \oplus \\
{\rm Hom}(\mathcal{O},\,\mathcal{T}(-\mathfrak{M})) & \oplus & {\rm
End}(\mathcal{T}(-\mathfrak{M}))
\end{array}
\simeq\mathcal{O}\oplus\mathcal{O}\oplus\mathcal{O}(n-2),$$ hence,
${\rm Aut}(\widetilde{\mathcal{L}_{\Theta_n}};\,l^+_1,\ldots,l^+_n)
  \simeq\mathbb{C}^*$.

As we have seen, the space of sections of the sheaf
$\mathcal{E}_{\Theta_n}$ on ${\bf\Theta}_n$ has dimension $(n-3)$;
hence,
$${\rm dim}\,\Gamma({\bf\Theta}_n,\,\mathcal{E}_{\Theta_n})+
{\rm dim}\,{\bf\Theta}_n=2\cdot(n-3),$$ that is, exactly the
dimension of the moduli space $\mathcal{M}'_n(2)$. Take a collection
of distinct points $\{x_1,\ldots,x_{n-3}\}\subset\mathbb{P}^1$ and a
collection of one-dimensional subspaces
$p_i\subset\widetilde{\mathcal{L}_{\Theta_n}}|_{x_i},\,i=1,\ldots,n-3$
and perform the modifications
$$A:=(x_1,p_1)^{\rm up}\circ\ldots\circ(x_{n-3},p_{n-3})^{\rm up}\,:\quad
\widetilde{\mathcal{L}_{\Theta_n}}\longrightarrow
\widetilde{\mathcal{L}},$$ where $\widetilde{\mathcal{L}}$ is a rank
2 bundle of degree -1 on $\mathbb{P}^1$. As it was shown, this gives
us a map from $\mathcal{M}'_n(2)$ to the symmetric product $({\rm
Tot}(\mathbb{P}^1,\Omega(\mathfrak{M})))^{(n-3)}$ at the generic
point.

Next, choose the unique connection
$\nabla_{\Theta_n}(p_1,\ldots,p_{n-3})\in\mathcal{E}_{\Theta_n}$
such that the subspaces $p_1,\ldots,p_{n-3}$ are invariant for it.
The modification of the connection is
$$A:\quad\nabla_{\Theta_n}(p_1,\ldots,p_{n-3})\longrightarrow
\widetilde{\nabla}=\nabla_{\Theta_n}(p_1,\ldots,p_{n-3})-
\sum\limits_{i=1}^{n-3}{\bf P}_{p_i}\frac{dz}{z-x_i},$$ where ${\bf
P}_{p_i}$ are the projectors on the (invariant) one-dimensional
subspaces $p_1,\ldots,p_{n-3}$. Note that this correspondence is
isomonodromic and the terms ${\displaystyle {\bf
P}_{p_i}\frac{dz}{z-x_i}}$ does not change the monodromy of the
connection and the points $x_1,\ldots,x_{n-3}$ are called
\emph{apparent singularities} of the connection
$\widetilde{\nabla}$. Originally, the apparent singularities were
introduced in \cite{F} by L. Fuchs; more detailed approach to the
Fuchsian equations and systems one can find in the books \cite{B},
and \cite{AB}.

In this way, we interpret the Hecke correspondence between the
moduli spaces ${\bf\Theta}_n$ and $\mathcal{M}'_n(2)$ as the
deformation of the most degenerate locus ${\bf\Theta}_n$ of $D$ in
the fibred space ${\rm Tot}({\bf\Theta}_n,\,\mathcal{E}_{\Theta_n})$
performed by modifications of the connection $\nabla_{\Theta_n}$
with apparent singularities ${\bf
P}_{p_i}{\displaystyle\frac{dz}{z-x_i}}$. In the case when $x_i\in
S$, the dynamics of the isomonodromic system becomes discrete and
presented by the lattice $C_n$; for calculations see the proposition
in Section 2; for applications to the relations between the special
functions, -- solutions of the Fuchsian equations, -- see the paper
\cite{O}.

\section{An example: the Painlev\'e-VI system}

\noindent Now, we illustrate our constructions of the \'etale
coordinates on the initial data space and its compactification in
the simplest example of the $sl(2)$-isomonodromic system with four
marked points called the sixth Painlev\'e system. In this section we
suppose that $\mathcal{L}$ is a rank 2 vector bundle on
$\mathbb{P}^1$ with ${\rm Det}\mathcal{L}\simeq\mathcal{O}$ and a
logarithmic connection $\nabla$ with eigenvalues $(\lambda_i,
-\lambda_i)$ of the residues at four singularities $a_i,\,i=1,..,4$.
So we have a modulus $\mathfrak{M}=\sum a_i$ and modulo projective
transformations of $\mathbb{P}^1$ by the three-dimensional group
$PGL(2,\mathbb{C})$ we can suppose $\mathfrak{M}=0+1+t+\infty$,
where $t:=r(a_1,a_2,a_3,a_4)$ is the cross-ratio; however,
$\nabla:\mathcal{L}\rightarrow\mathcal{L}\otimes\Omega^1(\mathfrak{M})$.

Following the ideas of previous sections, we shall investigate the
geometry of the moduli space $\mathcal{M}_4$ of such pairs
$(\mathcal{L},\nabla)$; its biggest cell is isomorphic to the
symplectic quotient $\EuScript{O}_1\times\ldots\times
\EuScript{O}_4//SL(2,\mathbb{C})$. We identify it with the phase
space of the Schlesinger system with four points on $\mathbb{P}^1$,
called the sixth Painlev\'e equation. We define suitable coordinates
using the geometric construction of the Schlesinger system from
\cite{AL}. Then we construct a natural compactification of the phase
space also considered in \cite{AL}, which is coincide with the
Okamoto compactification constructed in \cite{Oka}. At the end, we
discuss the geometric realization of the dynamics and the interplay
with the apparent singularities which is original.

First, consider the configuration space of the Painlev\'e-VI system.
It is the moduli space of so-called quasi-parabolic bundles
$\mathcal{N}_4$. Precisely, $\mathcal{N}_4$ is the moduli space of
the collections
$$(\mathcal{L};\quad\phi:{\rm Det}\mathcal{L}\simeq\mathcal{O};\quad
l_1,\ldots ,l_4),$$ where $\mathcal{L}$ is a rank 2 bundle with a
horizontal isomorphism $\phi$ and $l_i\subset\mathcal{L}|_{a_i}$
are one-dimensional subspaces; there is a canonical surjection
$\pi:\,\mathcal{M}_4\twoheadrightarrow\mathcal{N}_4$ defined by
$$(\mathcal{L},\nabla;\,\lambda _1,\ldots ,\lambda_n)\mapsto
(\mathcal{L};\, l_i^+:={\rm ker}({\rm Res}_{a_i}\nabla-\lambda
_i),\,i=1,\ldots ,4).$$ In fact, the configuration space
$\mathcal{N}_4(2)$ is parameterized by the $x$ coordinate. As we
have seen each pair $x_i,p_i$ naturally parameterize the non-trivial
bundle ${\rm Tot}(\mathbb{P}^1,\Omega(\mathfrak{M}))$; in this way
it is interesting to calculate the map $\pi$.

\subsection{Geometry of $\mathcal{N}_4(2)$}

\noindent Let us describe the configuration space of four
eigenvectors in the two-dimensional vector space or the
configurations of four points $l_1,\,l_2,\,l_3,\,l_4$ in
$\mathbb{P}^1$ modulo the action of $PGL(2)$. In our description we
follow Mumford's approach (see \cite{MS}).

The invariant of the configuration is the cross-ratio
$$r(l_1,l_2,l_3,l_4):=\frac{l_1-l_3}{l_1-l_4}\cdot\frac{l_2-l_4}{l_2-l_3};$$
naturally, it is a coordinate on $\mathcal{N}_4(2)$. Since we have
the action of the projective group $PGL(2,\mathbb{C})$ we can
suppose
$$l_1=X,\quad l_2=1,\quad l_3=0,\quad
l_4=\infty,\quad\mbox{hence,}\quad r(l_1,l_2,l_3,l_4)=X;$$ let us
calculate the behavior of $X=r(l_1,l_2,l_3,l_4)$ under the action of
the permutational factor-group
$$0\longrightarrow(\mathbb{Z}/2\mathbb{Z})^2\longrightarrow\mathfrak{S}_4
\longrightarrow\mathfrak{S}_3\longrightarrow 1.$$

The possible values of the cross-ratio are
$1-X,\,X^{-1},1-X^{-1}$. For example the value
$$1-X=1-\frac{l_1-l_3}{l_1-l_4}\cdot\frac{l_2-l_4}{l_2-l_3}=
\frac{l_4-l_3}{l_4-l_1}\cdot\frac{l_2-l_1}{l_2-l_3}$$ corresponds to
two different permutations: $(14):=l_1\leftrightarrow l_4$ and
$(23):=l_2\leftrightarrow l_3$. Thus, it corresponds to two
different quasi-parabolic bundles: one with $\{l_4=l_1\neq l_2\neq
l_3\neq l_1\}$ and another with $\{l_3=l_2\neq l_1\neq l_4\neq
l_2\}$. In this way if the two of the four points on the Riemann
sphere try to glue, then two others glue too: $X\rightarrow\infty$
if and only if $1\rightarrow 0$. Moreover, for each value
$X=0,\,X=1,\,X=\infty$, there are two different configurations of
quasi-parabolic bundles. Note that the configuration of the
quasi-parabolic bundle for the value
$X=r(l_1,l_2,l_3,l_4)=t=r(a_1,a_2,a_3,a_4)$ corresponds to the
nontrivial bundle
$\mathcal{L}\simeq\mathcal{O}(1)\oplus\mathcal{O}(-1)$.

Choose a basis in the two-dimensional fiber of our bundle:
$\mathcal{L}|_{a_i}:=\langle l_2,l_3\rangle$; then
$$\left\{
\begin{array}{lcl}
l_1=\alpha\cdot l_2+\beta\cdot l_3 =l_2+l_3; \\
l_2=1\cdot l_2+0\cdot l_3; \\
l_3=0\cdot l_2+1\cdot l_3; \\
l_4=\gamma\cdot l_2+\delta\cdot l_3 =l_2+r(\alpha,\beta,\gamma,\delta)
\cdot l_3
\end{array}\right. ,\qquad X=r(\alpha,\beta,\gamma,\delta);$$
consider the action of pairs of modifications on our bundle (see
Section 2):
$$(a_2,l_2)^{up}:\,\mathcal{L}\rightarrow\mathcal{L}',\quad
\langle l_2,l_3\rangle\rightarrow\langle
l_2':=\frac{l_2}{X-a_2},\,l_3\rangle,$$
$$(a_3,l_3)^{low}:\,\mathcal{L}'\rightarrow\widetilde{\mathcal{L}},\quad
\langle l'_2,l_3\rangle\rightarrow\langle\widetilde{l_2}:=
\frac{X-a_3}{X-a_2}\cdot l_2,\,l_3\rangle.$$ We have the modified
eigenvectors
$$\left\{
\begin{array}{lcl}
\widetilde{l_1}=\left({\displaystyle
\frac{X-a_3}{x-a_2}\cdot l_2+l_3}\right)_{X=a_1}=l_2+l_3; \\
\\
l_2=1\cdot l_2+0\cdot l_3; \\
\\
l_3=0\cdot l_2+1\cdot l_3; \\
\\
\widetilde{l_4}=\left({\displaystyle
\frac{X-a_3}{X-a_2}\cdot l_2+
r(\alpha,\beta,\gamma,\delta)\cdot l_3
}\right)_{X=a_4}=
r(a_1,a_2,a_3,a_4)\cdot l_2+r(\alpha,\beta,\gamma,\delta)\cdot l_3
\end{array}\right.;$$
if $r(\alpha,\beta,\gamma,\delta)\rightarrow t=r(a_1,a_2,a_3,a_4)$,
then $\widetilde{l_4}\rightarrow\widetilde{l_1}$. An analogous
calculation with the pair of modifications $(a_1,l_1)^{\rm
up}(a_4,l_4)^{\rm low}$ shows that the case
$\widetilde{l_2}\rightarrow\widetilde{l_3}$ gives the same value
$x=t$, hence, this value corresponds to two different nontrivial
quasi-parabolic bundles, and finally we have the following\\
{\bf Statement.} (\cite{AL}) $\mathcal{N}_4$ is isomorphic to
two copies of $\mathbb{P}^1$ glued outside $\{0,1,t,\infty\}$.\\
The action of the pairs of modifications on $\mathcal{N}_4$ is
evident and it presents the affine $\widehat{D}_4$ lattice.

\subsection{Geometry of $\mathcal{M}_4(2)$}

\noindent

Describe the geometry of the moduli space of the collections
$$(\mathcal{L},\nabla;\phi:{\rm Det}\mathcal{L}\simeq\mathcal{O};
\lambda_1,\lambda_2,\lambda_3,\lambda_4),$$ where $\mathcal{L}$ is a
rank 2 vector bundle with fixed holomorphic structure $\phi$ on the
determinant, and $\nabla$ is a logarithmic connection with fixed
eigenvalues of the residues at the points of the support $S$ of the
modulus $\mathfrak{M}=0+1+t+\infty$ on $\mathbb{P}^1$. Put the
eigenvalue condition
$$\sum\epsilon_i\lambda_i\notin\mathbb{Z},\qquad
(\epsilon_1, \epsilon_2, \epsilon_3, \epsilon_4)\in(\mathbb{Z}/{\rm
2}\mathbb{Z})^4$$ which provides the irreducibility of our pair
$(\mathcal{L},\nabla)$. Our notion of stability (see 3.1) of our
pair $(\mathcal{L},\nabla)$ implies that neither of the eigenvectors
$l_i^+:={\rm ker}({\rm Res}_{x_i}\nabla-\lambda _i)$ may lie in the
sub-bundle $\mathcal{L}_0\simeq\mathcal{O}(1)$. Modify our bundle,
say, at $(\infty,l_\infty^+)^{\rm low}$, we necessarily get the
bundle
$\widetilde{\mathcal{L}}\simeq\mathcal{O}\oplus\mathcal{O}(-\infty)$;
this modification presents an isomorphism of $\mathcal{M}_4$ with
$\mathcal{M}_4'$, which is the moduli space of the following
collections.
$$(\widetilde{\mathcal{L}},\widetilde{\nabla};\quad\widetilde{\phi}:
{\rm
Det}\widetilde{\mathcal{L}}\simeq\mathcal{O}(-\infty);\quad(\lambda_1,
-\lambda_1),\ldots , (\lambda_\infty, 1-\lambda_\infty)).$$ In this
way, we get a uniquely defined sub-bundle
$$\widetilde{\mathcal{L}}\supset\widetilde{\mathcal{L}_0}\simeq\mathcal{O}$$
with the standard connection $d$. Restrict our connection to the
sub-bundle and consider the operator
$$A(z):={\rm Id}\oplus(\nabla|_{\widetilde{\mathcal{L}_0}}-\partial_z):
\quad\mathcal{O}\oplus\widetilde{\mathcal{L}_0}\longrightarrow
\widetilde{\mathcal{L}}\otimes\Omega^1(\mathfrak{M}).$$  Our pair is
irreducible, ${\rm
Im}(\nabla|_{\widetilde{\mathcal{L}_0}}-\partial_z)
(\widetilde{\mathcal{L}_0}) \nsubseteq\widetilde{\mathcal{L}_0},$
hence,
$$A(z):={\rm Id}\oplus(\nabla|_{\widetilde{\mathcal{L}_0}}-\partial_z):
\quad\mathcal{O}\oplus\mathcal{T}(-\mathfrak{M})\longrightarrow
\widetilde{\mathcal{L}}.$$ The determinant ${\rm Det}A(z)$ has a
simple pole at some point $x$ and, moreover, $A(z)=(x,p)^{\rm up}$;
the variables $x$ and $p$ are the canonical coordinates on the
two-dimensional initial data space $\mathcal{M}_4$ of our
isomonodromic system. The surface $\mathcal{M}_4$ is noncompact and
has a structure of a fibred space over $\mathcal{N}_4$. Note that in
our case the cohomological calculations are very simple:
$\mathcal{E}\simeq\mathcal{O}(-2)^*\otimes\mathcal{O}(-1)\otimes
\Omega\simeq\mathcal{O}(-1)$ and $H^1(\mathcal{E})=0$, hence,
$\mathcal{M}_4\simeq K_4'$

\subsection{Geometry of the Painlev\'e-VI system}
\noindent
As we have seen the moduli space $\mathcal{M}'_4(2)$ is
the non-compact surface\\

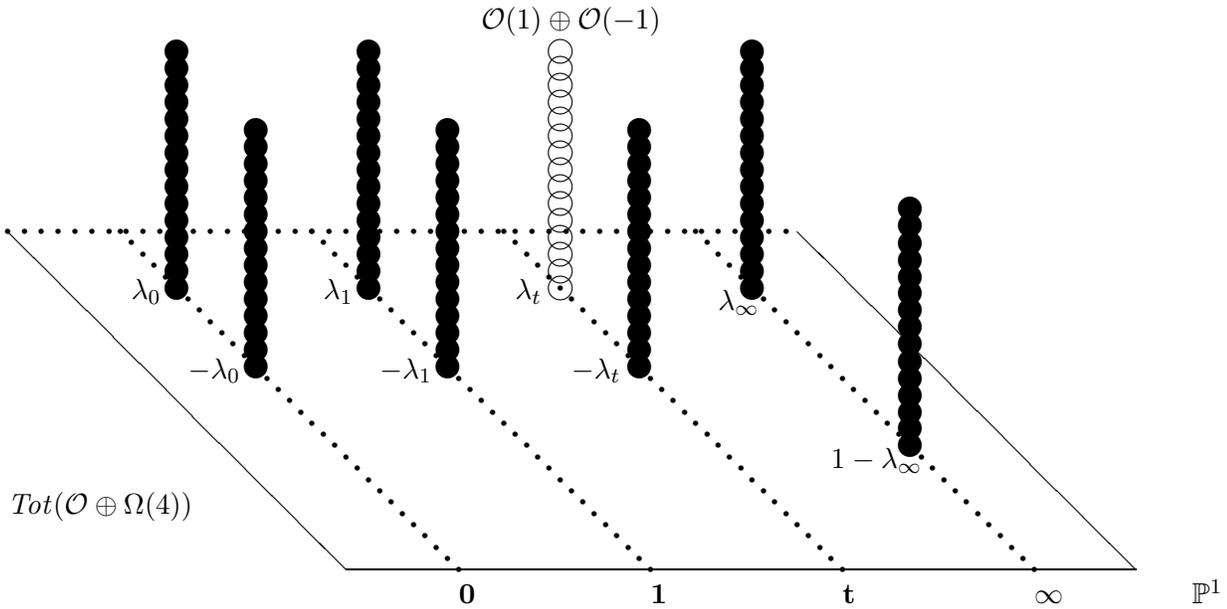
\begin{picture}(100,70)
\put(-10,40){\line(1,-1){30}}
\put(20,10){\line(1,0){70}}
\put(90,10){\line(-1,1){30}}
\multiput(-10,40)(1.5,0){47}%
{\circle*{0.5}}

\put(95,7){$\mathbb{P}^1$}
\put(30,7){{\bf 0}}
\put(47,7){{\bf 1}}
\put(64,7){{\bf t}}
\put(81,7){$\infty$}
\put(-10,15){${\it Tot}(\mathcal{O}\oplus\Omega(4))$}

\multiput(0,40)(1,-1){31}%
{\circle*{0.5}}
\multiput(17,40)(1,-1){31}%
{\circle*{0.5}}
\multiput(34,40)(1,-1){31}%
{\circle*{0.5}}
\multiput(51,40)(1,-1){31}%
{\circle*{0.5}}

\multiput(5,35)(0,1.5){15}%
{\circle*{2}}
\multiput(22,35)(0,1.5){15}%
{\circle*{2}}
\multiput(39,35)(0,1.5){15}%
{\circle{2}}
\put(32,58){$\mathcal{O}(1)\oplus\mathcal{O}(-1)$}
\multiput(56,35)(0,1.5){15}%
{\circle*{2}}

\multiput(12,28)(0,1.5){15}%
{\circle*{2}}
\multiput(29,28)(0,1.5){15}%
{\circle*{2}}
\multiput(46,28)(0,1.5){15}%
{\circle*{2}}
\multiput(70,21)(0,1.5){15}%
{\circle*{2}}

\put(1,34){$\lambda_0$}
\put(18,34){$\lambda_1$}
\put(35,34){$\lambda_t$}
\put(53,33){$\lambda_\infty$}

\put(6,27){$-\lambda_0$}
\put(23,27){$-\lambda_1$}
\put(40,27){$-\lambda_t$}
\put(63,19){$1-\lambda_\infty$}
\end{picture}\\
The exceptional divisor at a point $(t,\lambda_t)$ corresponds to
the collection
$(\widetilde{\mathcal{L}},\widetilde{\nabla};\quad\widetilde{\phi}:
{\rm Det}\widetilde{\mathcal{L}}\simeq\mathcal{O}(-\infty);\quad
(\lambda_1, -\lambda_1),\ldots , (\lambda_\infty,
1-\lambda_\infty))$ with a nontrivial bundle
$\widetilde{\mathcal{L}}\simeq\mathcal{O}(1)\oplus\mathcal{O}(-1)$.

In this way we have the following presentation of the initial data
space
$$\mathcal{M}_4(2)\simeq K'_4:=({\rm Bl}_{R^{-1}(\lambda_i^\pm)}
{\rm Tot}
(\mathbb{P}^1,\,\mathcal{O}(2)))\setminus\bigsqcup\widetilde{F_i},\quad
i=1,\ldots;4$$ it is isomorphic to the moduli space of the stable
FH-sheaves
$$(\mathcal{O}\oplus\mathcal{T}(-0-1-t-\infty)\subset\mathcal{O}\oplus
\mathcal{O}(-\infty))$$ of level 1. In other words, the coordinates
$(x,p)$ on the initial data space present it as the moduli space of
exact sequences
$$0\longrightarrow\mathcal{O}\oplus\mathcal{T}(-4)\longrightarrow
\mathcal{O}\oplus\mathcal{O}(-\infty)\longrightarrow
\delta_x\otimes p\otimes\mathcal{T}_x\longrightarrow 0$$ such that
$p<\infty$ and if $x=a\in S$ then $p=\lambda_a ^\pm$.

Consider the natural symplectic form $\varpi=dx\wedge dp$ on
$\mathbb{P}(\mathcal{O}\oplus\Omega(4))$, and let us look at its
behavior when $x\in S$. At singular points of the connection the
dynamics is discrete and performed by the lattice $\widehat{F}_4$.
We blow-up eight points $(x,p)=(a,\lambda_a^\pm),\,a\in S$, on the
surface $\mathbb{P}(\mathcal{O}\oplus\Omega(4))$; locally this
procedure performed by $p=s\cdot x$ for $s$ a coordinate on the
exceptional divisor. Then, remove four fibers
$\widetilde{F_a}:=\{a,p\}\subset\mathbb{P}(\mathcal{O}\oplus\Omega(4))$
and in this way at $x=a$ we have two exceptional curves with
$$ds=\frac{dp}{x}-s\cdot\frac{dx}{x}.$$

The compactifying set is exactly the divisor of poles of the
symplectic form $\varpi=dx\wedge dp$, and it performs the
degeneration of an elliptic curve $C$. The divisor is\\
\begin{picture}(100,22)
\put(0,14){$D=(2\cdot s_\infty+
\widetilde{F_0}+\widetilde{F_1}+\widetilde{F_t}+
\widetilde{F_\infty})^{\rm red}=$}
\multiput(50,18)(0.5,0){60}{\circle*{1}}
\multiput(52,18)(0.5,-0.5){25}{\circle*{1}} \put(63,3){{\bf 0}}
\multiput(60,18)(0.5,-0.5){25}{\circle*{1}} \put(71,3){{\bf 1}}
\multiput(68,18)(0.5,-0.5){25}{\circle*{1}} \put(79,3){{\bf t}}
\multiput(76,18)(0.5,-0.5){25}{\circle*{1}} \put(87,3){$\infty$}
\end{picture}\\
it is defined by the conditions $p=\infty$ and $l_a ^+=l_a
^-,\,a=0,1,t,\infty$. Let $\widetilde{\mathcal{L}}$ be the bundle
corresponding to a point on the compactifying divisor and perform
the lower modification, say, at $a=0$ in the direction
$$l_0^+\subset\mathcal{O}|_{z=0}\subset(\mathcal{O}\oplus
\mathcal{O}(-1))|_{z=0}.$$ We get the bundle
$\widetilde{\mathcal{L}_D}\simeq\mathcal{O}\oplus\mathcal{T}(-4)$,
and we have an isomorphism of $D$ with the moduli space of the
collections
$$(\widetilde{\mathcal{L}_D},\,\nabla_D,\,
\phi',\,(\widetilde{\lambda_i^+},\widetilde{\lambda_i^-})),$$ where
$\widetilde{\mathcal{L}_D}$ is a bundle of degree $-2$ on
$\mathbb{P}^1$ with the horizontal isomorphism $\phi':{\rm
Det}\widetilde{\mathcal{L}}'\widetilde{\rightarrow}\mathcal{O}(-0-\infty)$
and the connection $\nabla_D$ with the following eigenvalues of
residues
$(\widetilde{\lambda_0^+},\widetilde{\lambda_0^-})=(\lambda_0,1-\lambda_0),$
$$(\widetilde{\lambda_1^+},\widetilde{\lambda_1^-})=(\lambda_1,-\lambda_1),
\quad
(\widetilde{\lambda_t^+},\widetilde{\lambda_t^-})=(\lambda_t,-\lambda_t),
\quad
(\widetilde{\lambda_\infty^+},\widetilde{\lambda_\infty^-})=
(\lambda_\infty,1-\lambda_\infty).$$

Finally we have the following diagram
$$\mathcal{O}\oplus\mathcal{T}(-4)\quad
\stackrel{(x,p)^{\rm up}}{\longrightarrow}\quad
\mathcal{O}\oplus\mathcal{O}(-1)\quad
{\longrightarrow\atop\longleftarrow}\quad \left[\begin{array}{l}
\mathcal{O}\oplus\mathcal{O}\\
\mathcal{O}(1)\oplus\mathcal{O}(-1)
\end{array}\right..$$
The right two arrows ${\rightarrow\atop\leftarrow}$ denote the
action of discrete $\widehat{F}_4$-symmetries (see \cite{AL},
\cite{O}) and the left arrow $\stackrel{(x,p)^{\rm
up}}{\longrightarrow}$ in terms of the connections is
$$(x,p)^{\rm up}:\widetilde{\nabla}=\nabla_D(p)-
{\bf P}_p{\displaystyle\frac{dz}{z-x}}.$$ Note here that the
connection $\nabla_D$ is not uniquely defined. Such connections on
the bundle $\mathcal{O}\oplus\mathcal{O}(-2)$ form a
one-dimensional affine space and we choose uniquely the connection
$\nabla_D(p)$ for which the direction $p$ is proper; otherwise, as
it was shown we can get the quadratic pole of $\widetilde{\nabla}$
at $z=x$.

The term ${\bf P}_p{\displaystyle\frac{dz}{z-x}}$ does not change
the monodromy of connections and the simple pole at $z=x$ is an
apparent singular point for the appropriate Fuchsian system. In this
way we perform the isomonodromic system Painlev\'e-VI as the
deformation of the moduli space $D$ by the Hecke correspondence
$(x,p)^{\rm up}$.

For the interpretation of the Painlev\'e-VI system as a deformation
of the compactifying divisor in terms of the Kodaira-Spencer theory
see \cite{T}.

\end{document}